\documentclass[10pt,twoside]{article}%{ % Specifies the document style.

\usepackage{times,fancyhdr}
\usepackage{cite}
\usepackage{graphicx}
\usepackage{amsmath,amsfonts, amssymb}

\usepackage[ansinew]{inputenc}

\usepackage{latexsym}

\def\a{\alpha}
\def\b{\beta}
\def\g{\gamma}
\def\d{\delta}

\def\vt{\vartheta}

\def\a{\alpha}
\def\b{\beta}
\def\m{\mu}

\def\g{\gamma}
\def\d{\delta}
\def\g{\gamma}

\def\r{\rho}

\def\L{{\mathcal{L}}}
\def\T{{\mathcal{T}}}

\def\C{{\mathcal{C}}}
\def\F{{\mathcal{F}}}
 
 \def\M{{\mathcal{M}}}
 \def\Q{{\mathcal{Q}}}
  \def\A{{\mathcal{A}}}
 \def\B{{\mathcal{B}}}
 \def\C{{\mathcal{C}}}
 \def\A{{\mathcal{A}}}
  \def\R{{\mathcal{R}}}
  
  \def\GG{{\it{\Gamma}}}
  \def\hi{{\hat{i}}}
  \def\hj{{\hat{j}}}
\def\G{\Gamma}
\def\oG{\stackrel{{\rm o}}{\Gamma}{\!}}
 \def\oGG{\stackrel{{\rm o}}{\it \Gamma}{\!}}
\def\LG{\stackrel{{\rm *}}{\Gamma}{\!}}

\def\oTT{\stackrel{{\rm o}}{\T}{\!}}
\def\brr{\begin{eqnarray}}
\def\err{\end{eqnarray}}
\def\brn{\begin{eqnarray*}}
\def\ern{\end{eqnarray*}}
\date{}

\title{Coframe geometry and gravity} % Declares the document's title.
\author{Yakov Itin \\
Institute of Mathematics, Hebrew University of
 Jerusalem \\ and Jerusalem College of Technology,\\ Jerusalem 91904,
 Israel, \\email: {\it itin@math.huji.ac.il}}

\begin{document} % End of preamble and beginning of text.
\pagestyle{fancy}
\fancyhead{} % clear all header fields
\fancyhead[EC]{Yakov Itin}
\fancyhead[EL,OR]{\thepage}
\fancyhead[OC]{Coframe geometry and gravity}
\fancyfoot{} % clear all footer fields
\renewcommand\headrulewidth{0.5pt}
\addtolength{\headheight}{2pt} % make space for the rule

\numberwithin{equation}{section}
\maketitle % Produces the title.

\begin{abstract}
 The possible extensions of GR for  description of fermions on a curved space, for supergravity and for loop quantum gravity require a richer set of 16 independent variables. These variables can be assembled in a coframe field, i.e., a local set of four linearly independent 1-forms.  In this chapter we study  the gravity field models based on a coframe variable alone. 
 We give a short review of the  coframe gravity. This model  has the viable Schwarzschild  solutions even being alternative to the standard GR. Moreover, the coframe model treating of the gravity energy may be preferable to the ordinary GR where the gravity energy cannot be defined at all. A principle problem that the coframe gravity  does not have any connection to a specific geometry even being constructed from the geometrical meaningful objects. A geometrization of the coframe gravity is an aim of this chapter. 
 We construct a complete class of the coframe connections which are  linear in the first order
 derivatives of the coframe field on an $n$ dimensional manifolds with and without a metric.
  The subclasses of the torsion-free, metric-compatible and flat
  connections are derived. We also study the behavior of the
 geometrical structures under local transformations of the coframe. The remarkable fact
 is an existence of  a subclass of connections which are invariant when the infinitesimal transformations satisfy the Maxwell-like system of equations. 
In the framework of the  coframe geometry construction,  we propose a geometrical action for the coframe gravity.  It is similar to the Einstein-Hilbert action of GR, but the scalar curvature is constructed from the general coframe connection. We show that this geometric Lagrangian is equivalent to the coframe Lagrangian up to a total derivative term. Moreover there is a family of coframe connections which Lagrangian does not include the higher order terms at all. In this case, the equivalence is complete. 

\end{abstract}
%\pacs{03.30.+p, 11.30.Cp, 42.25.Lc}

\date{\today}

% I ask the editors to review the English for style. 
\maketitle
\newpage
\tableofcontents

%--------------------------------------------
\section{Introduction. Why  do we have to go beyond  Riemannian geometry?}
%--------------------------------------------
 General relativity (GR) is, probably, the best of the known theories
 of gravity.
 From mathematical and aesthetic points of view, it
 can be used as a standard of what a physical theory has to be. 
 Up to this day, the Einstein theory  is in a very  good agreement
 with the observation data. 
Probably the main idea of Einstein's GR is that the physical
properties of the gravitational field  are in one-to-one
correspondence with the geometry of the base manifold. The 
standard GR is based on  a Riemannian geometry with a unique metric tensor 
and a unique Levi-Civita connection constructed from this tensor.
Hence, the gravity field equations of GR  predicts a unique (up to
diffeomorphism transformations)  metric tensor and consequently a unique geometry. 
Therefore  any  physical field except of gravity can not have an intrinsic geometrical sense in 
the Riemannian geometry. 

After the classical works of Weyl, Cartan and others, we know that the Riemannian construction is not a unique possible geometry.  
A most general geometric framework involves independent metric and independent connection. 
A gravity field model based on this general geometry (Metric-affine gravity) was studied intensively, see \cite{Hehl:1976kj}---\cite{Obukhov:2006gy} and the references given therein. Probably a main problem of this construction is a huge number of geometrical fields which do not find their physical partner. 

In this chapter we study a much more economical construction based on a unique geometrical object --- coframe field.  Absolute (teleparallel) frame/coframe variables (rep\`er, vierbein, ...) were
 introduced in physics by Einstein in 1928 with an aim of a unification 
 of gravitational and electromagnetic fields (for classical references, 
 see \cite{cit1}). The physical models for gravity based on the coframe variable are well studied, see  \cite{Hayash:1979}---\cite{Nashed:2007sc}. In some aspects such models are even preferable from the standard GR. In particular, they involve a meaningful definition of the gravitational energy, which is in a proper correspondence with the Noether procedure. 
Moreover some problems inside and beyond Einstein's gravity
 require a richer set of 16 independent variables of the coframe. 
In the following issues of gravity, the coframe is not
only a useful tool but often it cannot even be replaced
by the standard metric variable: 
(i) Hamiltonian formulation \cite{Ashtekar:1987gu},\cite{Deser:1976ay}; 
(ii) positive energy proofs \cite{Nester:1994du}; 
(iii) fermions on a curved manifold \cite{Deser:1974cy},\cite{spin}; 
(iv) supergravity \cite{VanNieuwenhuizen:1981ae}; 
(v) loop quantum gravity  \cite{Perez:15jz}.

 Unfortunately, in the coframe gravity models, the proper connection between physics and  the underlying geometry is lost. In this chapter, we propose a way of geometrization of the coframe gravity.  In particular, we study which geometric structure can be constructed 
 from the  vierbein  (frame/coframe) variables  and which gravity field models 
 can be related to this geometry. 
 
The organization of the  chapter is as follows: 

In the first section, we give a brief account of the gravity field model based on the coframe field instead of the pure metrical construction of GR. We discuss the following features:  
(i) The coframe gravity is described by a 3-parametric 
set of models; 
(ii)  All the coframe models are derivable from a Yang-Mills-type Lagrangian; 
(iii) The coframe field equations are well defined  for all values of the parameters. 
Only for the pure GR case, the system id degenerated to 10 equations for 16 variables;
(iv) The energy-momentum tensor of the coframe field is well defined for all models except GR. In the latter case the tensor nature of the energy-momentum expression is lost; 
(v) There is a subset of viable fields with a unique spherical symmetric solution, which corresponds to  Schwarzschild metric; 
(vi) The same subset is derived by the requirement of the free field limit approximation. 
All these positive properties make the coframe gravity a relevant subject of investigation.

In section 2, we construct a geometrical structure based on a coframe variable as unique building block. In an addition to the coframe volume element and  metric, we present a most general coframe connection. The Levi-Civita and flat connections are special cases of it. The  torsion and nonmetricity tensors of the general coframe connection are calculated. We identify the subclasses of symmetric (torsion-free) connections and of metric-compatible connections. The unique symmetric metric-compatible connection is of Levi-Civita. We study the transformations of the coframe field and identify a subclass of connections which are invariant under restricted coframe transformations. Quite remarkable that restriction conditions are approximated by a Maxwell-type system.

In section 3, we are looking for a geometric representation of the gravity coframe model. 
The main result is that the free-parametric gravity coframe Lagrangian can be replaced 
by a standard Einstein-Hilbert Lagrangian, when the curvature scalar is calculated on a general coframe connection. The standard GR Lagrangian contains a second order derivative term which appears in the form of the total derivative. This term does not influence  the field equation, but it cannot be consistently removed. We show that there is a set of 
 coframe connections which Einstein-Hilbert Lagrangian does not involve the second order derivative term at all. 

In the last section, some proposals  of  possible developments of a geometrical coframe construction and its applications to gravity are presented.  

%-----------------------------
\section{Coframe gravity}
%-----------------------------
Let us give a brief account of gravity field models based on a coframe field. We refer to such models as {\it coframe gravity}.  
This is instead of the Einsteinian {\it metric gravity} based on a metric tensor field. 
 We will use here mostly the notations accepted in \cite{Itin:2001bp}. 
\subsection{Coframe Lagrangian}
Consider a smooth, non-degenerated coframe field $\{\vt^\a, \ \a=0,1,2,3\}$ defined 
on a $4D$  smooth differential manifold $M$.
The 1-forms $\vt^\a$ are declared to be pseudo-orthonormal.
Thus  a metric on $M$ is defined by 
\begin{equation}\label{3.1} 
g=\eta_{\a\b}\vt^\a\otimes\vt^\b\,,\qquad \eta_{\a\b}=(-1,1,1,1)\,.
\end{equation}
So, the coframe field $\vt^\a$ is considered as
a basic dynamical variable  while the metric $g$ is  treated as 
a secondary structure.
 
The coframe field is defined only up to
{\it global pseudo-rotations}, i.e. $SO(1,3)$ transformations. 
Consequently, the truly dynamical variable is  an equivalence
class of coframes   $[\vt^\a]$, while 
the global pseudo-rotations produce  an  equivalence relation on this class. 
Hence, in addition to the invariance under the diffeomorphic transformations
of the manifold $M$, the basic geometric structure
has to be  global (rigid) $SO(1,3)$ invariant.

Gravity is described  by differential 
invariants of the coframe structure. 
There is an important distinction between the diffeomorphic invariants of 
the metric and of the coframe structures. 
Since the metric invariants of the first order are trivial, the metric structure admits diffeomorphic invariants only of the second order or greater. 
A unique invariant of the second order is the scalar curvature. 
This expression is well known to play the key 
role of an  integrand in the Einstein-Hilbert action. 
 The coframe structure admits diffeomorphic and rigid $SO(1,3)$ 
invariants even of the first order. 
A simple example is the expression $e_\a\rfloor d\vt^\a$, see Appendix for notations and basic definitions. 
The operators, which are diffeomorphic invariants and global 
covariants,  can contribute to a general coframe field equation.  
A rich class of such equations is constructed in \cite{Itin:1999zs}. 
A requirement of derivability of the field equations from a Lagrangian strictly 
restricts the variety of possible options. 

We restrict the consideration to  
odd, quadratic (in the first order derivatives of the coframe field 
$\vt^\a$),  diffeomorphic, and global $SO(1,3)$ invariant Lagrangians.  
A general Lagrangian of such a type is represented by a linear combination 
of three 4-forms which are referred to as the Weitzenb\"{o}ck   invariants.  
Consider the exterior differentials of the basis 1-forms $d\vt^\a$ and 
introduce the coefficients of their  expansion in the basis of 
even 2-forms $\vt^{\a\b}$ 
\begin{equation}\label{4.2} 
d\vt^\a=\vt^\a_{i,j}dx^i\wedge dx^j=\frac 12 {C^\a}_{\b\g}\vt^{\b\g}\,.
\end{equation}
We use here   the abbreviation 
$\vt^{\a\b\cdots}=\vt^\a\wedge \vt^\b\wedge \cdots$.
By definition, the coefficients ${C^\a}_{\b\g}$ are  antisymmetric,  
${C^\a}_{\b\g}=-{C^\a}_{\g\b}.$ 
Their explicit expression can be given by the differential form notations (see Appendix) 
\begin{equation}\label{4.3} 
{C^\a}_{\b\g}=e_\g\rfloor(e_\b\rfloor d\vt^\a)\,.
\end{equation}
The symmetric form of a general second order coframe  Lagrangian  is given by 
\cite{Muench:1998ay} 
\begin{equation}\label{3.2}         
{}^{(\tt cof)}L=
%\frac 1{\ell^2} L= 
\frac 1{2\ell^2}\sum_{i=1}^3 \rho_{i} \; {}^{(i)}L\,,
\end{equation}
where $\ell$ denotes the Planck length constant, while $\rho_i$ are dimensionless parameters. 
The partial Lagrangian expressions are 
\brr\label{3.3}   
{}^{(1)}L &=&d\vt^\a \wedge *d\vt_\a=\frac 12 C_{\a\b\g}C^{\a\b\g}*1\,,\\
\label{3.4}   
{}^{(2)}L &=&
\left(d\vt_\a \wedge \vt^\a \right) \wedge*\left(d\vt_\b\wedge\vt^\b\right)=
\frac 12 C_{\a\b\g}\left(C^{\a\b\g}+C^{\b\g\a}+C^{\g\a\b}\right)*1\,,\\
\label{3.5}     
{}^{(3)}L &=& 
(d\vt_\a \wedge\vt^\b ) \wedge *\left(d\vt_\b \wedge \vt^\a \right)=\frac 12 \left(C_{\a\b\g}C^{\a\b\g}-2{C^\a}_{\a\g}{C_\b}^{\b\g}\right)*1\,.
\err
The 1-forms $\vt^\a$ are assumed to carry the dimension of length, while 
the coefficients $\r_i$ are dimensionless. 
Hence the total Lagrangian ${}^{(\tt cof)}L$ is  dimensionless. 
In order to simplify the formulas below we will use the Lagrangian 
$L=\ell^2 {}^{(\tt cof)}L$ which dimension is   length square. 
In other worlds  the geometrized units system with $G=c=\hbar=1$ is applied. 

Every term of the Lagrangian (\ref{3.2}) is independent of  
a specific choice of a coordinate system and invariant under a global 
(rigid) $SO(1,3)$ transformation of the coframe field. 
Thus, different choices of the free parameters $\r_i$ yield 
different rigid $SO(1,3)$ and diffeomorphic invariant classical field models. 
Some of them are known to  be applicable for description  of  gravity.

Let us rewrite the  coframe Lagrangian 
in a compact form
\begin{equation}\label{4.5}  
{}^{(\tt cof)}L=\frac 1{4} C_{\a\b\g}C_{\a'\b'\g'}\lambda^{\a\b\g\a'\b'\g'}*1\,,
\end{equation}
where the  constant symbols
\brr\label{4.6}  
\lambda^{\a\b\g\a'\b'\g'}&=&(\rho_1+\rho_2+\rho_3)\eta^{\a\a'}\eta^{\b\b'}\eta^{\g\g'}+
\rho_2(\eta^{\a\b'}\eta^{\b\g'}\eta^{\g\a'}+\eta^{\a\g'}\eta^{\b\a'}\eta^{\g\b'})
\nonumber \\&&
-2\rho_3\eta^{\a\g}\eta^{\a'\g'}\eta^{\b\b'}
\err
are introduced.  
It can be checked, by straightforward calculation, that these $\lambda$-symbols 
are invariant under a 
transposition of the triplets of indices: 
\begin{equation}\label{4.7}
\lambda^{\a\b\g\a'\b'\g'}=\lambda^{\a'\b'\g'\a\b\g}\,.
\end{equation}
We also  introduce an abbreviated notation 
 \begin{equation}\label{4.8}
F^{\a\b\g}=\lambda^{\a\b\g\a'\b'\g'}C_{\a'\b'\g'}\,.
\end{equation}
The total Lagrangian (\ref{3.2}) reads now as
\begin{equation}\label{4.10}
{}^{(\tt cof)}L=\frac 1{4} C_{\a\b\g}F^{\a\b\g}*1\,.
\end{equation}
This form of the Lagrangian will be used in sequel 
for the variation procedure. 
The Lagrangian (\ref{4.10}) can also be rewritten in a component free 
notations. 
Define one-indexed 2-forms: a {\it field strength form}  
 \begin{equation}\label{4.16}
\C^\a:=\frac 12 C^{\a\b\g}\vt_{\b\g}=d\vt^\a\,,
\end{equation}
and a {\it conjugate field strength form} $\F^\a:=\frac 12 F^{\a\b\g}\vt_{\b\g}$ 
 \begin{equation}\label{4.16a}
\F^\a=(\rho_1+\rho_3)\C^\a+
\rho_2e^\a\rfloor(\vt^\mu\wedge\C_\mu)-\rho_3\vt^\a\wedge(e_\mu\rfloor \C^\mu)\,.
\end{equation}
Another form of $\F^\a$ can be given via the irreducible (under the 
Lorentz group) decomposition of the 2-form $\C^\a$ (see \cite{Hehl:1994ue}, \cite{McCrea:1992wa}). 
Write 
\begin{equation}\label{4.17a}
\C^\a={}^{(1)}\C^\a+{}^{(2)}\C^\a+{}^{(3)}\C^\a,
\end{equation}
where
\brr\label{4.17b}
%{}^{(1)}\C^\a=\C^a-{}^{(2)}\C^\a-{}^{(3)}\C^\a\,,\nonumber\\
{}^{(2)}\C^\a=\frac 13 \vt^\a\wedge(e_\mu\rfloor \C^\mu)\,,\qquad
{}^{(3)}\C^\a=\frac 13 e^\a\rfloor(\vt_\mu\wedge \C^\mu)\,,
\err
while ${}^{(1)}\C^\a$ is the remaining part. 
Substitute (\ref{4.17b}) into (\ref{4.16a}) to obtain 
\begin{equation}\label{4.17c}
\F^\a=(\rho_1+\rho_3){}^{(1)}\C^\a+(\rho_1-2\rho_3){}^{(2)}\C^\a+
(\rho_1+3\rho_2+\rho_3){}^{(3)}\C^\a.
\end{equation}
The coefficients in (\ref{4.17c}) coincide with those calculated in 
\cite{Muench:1998ay}. \\ 
The 2-forms  $\C^\a$ and $\F^\a$ do not depend on a choice of 
a coordinate system. They  change as 
vectors by global $SO(1,3)$ transformations of the coframe. 
Using (\ref{4.16}) the coframe Lagrangian can be rewritten  as
 \begin{equation}\label{4.17}
{}^{(\tt cof)}L=\frac 1{2} \C_\a\wedge *\F^\a\,.
 \end{equation}
Observe that the Lagrangian (\ref{4.17}) is of the same form as the 
standard electromagnetic Lagrangian 
% \begin{equation}\label{4.18a}
${}^{(\tt cof)}L=\frac 1{2} F\wedge*F.$
% \end{equation}
Observe, however, that the coframe Lagrangian involves the vector valued 2-forms of the 
field strength, while the  electromagnetic Lagrangian is constructed of the 
the scalar valued 2-forms. 
%%%%%%%%%%%%%%%%%%%%%%%%%%%%%%%%%%%%%%%%%%%%%%%%%%%%
 \subsection{Variation of the Lagrangian}                %%%%%%% 4
%%%%%%%%%%%%%%%%%%%%%%%%%%%%%%%%%%%%%%%%%%%%%%%%%%%%
The Lagrangian (\ref{4.17}) depends on the coframe field $\vt^a$ and on its first 
order derivatives only. 
Thus the first order variation formalism guarantee the corresponding 
Euler-Lagrange equation to be at most of the second order. 
Consider the variation of the coframe Lagrangian (\ref{4.10}) with respect to small independent variations of the 1-forms $\vt^\a$. 
The $\lambda$-symbols (\ref{4.6}) are constants and obey the 
symmetry property (\ref{4.7}). 
Thus 
\begin{equation}
C_{\a\b\g}\d F^{\a\b\g}=C_{\a\b\g}\lambda^{\a\b\g\a'\b'\g'}\d C_{\a'\b'\g'}=
\d C_{\a\b\g} F^{\a\b\g}\,.
\end{equation}
Consequently 
the variation of the Lagrangian  (\ref{4.10}) takes the form 
\begin{equation}\label{4.11}  
\d L=\frac 1{2}\d C_{\a\b\g}F^{\a\b\g}*1-L*\d(*1)\,.
\end{equation}
The variation of the volume element is 
\brn   %\label{4.12} 
\d(*1)=-\d(\vt^{0123})=-\d\vt^0\wedge \vt^{123}-\cdots=
-\d\vt^0\wedge *\vt^0-\cdots %\nonumber  \\
%\label{4.12a} 
%&=&
=\d \vt^\a\wedge*\vt_\a\,.
\ern   
Thus the second term of (\ref{4.11}) is given by 
\begin{equation}\label{4.12}
L*\d(*1)=(\d\vt^\a\wedge *\vt_\a)*L=-\d\vt^\a\wedge(e_\a\rfloor L)\,.
\end{equation}
As for the variation of the $C$-coefficients, we calculate them by equating  
the variations  of the two sides of the equation (\ref{4.2})
\begin{equation}
\d d\vt_{\a}=\frac 12 \d C_{\a\mu\nu}\vt^{\mu\nu}+C_{\a\mu\nu}\d\vt^{\mu}\wedge\vt^\nu\,.
\end{equation}
Use the formulas (\ref{a0-6}) and (\ref{a0-7}) to derive  
\brn
\d d\vt_{\a}\wedge*\vt_{\b\g}&=&
\frac 12 \d C_{\a\mu\nu}\vt^{\mu\nu}\wedge*\vt_{\b\g}+
C_{\a\mu\nu}\d\vt^{\mu}\wedge\vt^\nu\wedge*\vt_{\b\g}\\
%%
%&=&\frac 12 \d C_{amn}\vt^{m}\wedge *^2(\vt^n\wedge*\vt_{bc})+
%C_{amn}\d\vt^{m}\wedge*^2(\vt^n\wedge*\vt_{bc})\\
%%
&=&-\frac 12 \d C_{\a\mu\nu}\vt^{\mu}\wedge *(e^\nu\rfloor\vt_{\b\g})-
C_{\a\mu\nu}\d\vt^{\mu}\wedge*(e^\nu\rfloor\vt_{\b\g})\\
%%
%&=&-\frac 12 \d C_{amb}\vt^{m}\wedge *\vt_{c}+
%\frac 12 \d C_{amc}\vt^{m}\wedge *\vt_{b}\\
%&&-C_{amb}\d\vt^{m}\wedge*\vt_c+C_{amc}\d\vt^{m}\wedge*\vt_b\\
%%
%&=&\frac 12 \d C_{amb}*^2(\vt^{m}\wedge *\vt_{c})-
%\frac 12 \d C_{amc}*^2(\vt^{m}\wedge *\vt_{b})\\
%&&-C_{amb}\d\vt^{m}\wedge*\vt_c+C_{amc}\d\vt^{m}\wedge*\vt_b\\
%%
&=&\d C_{\a\b\g}*1-2\d\vt^{\mu}\wedge C_{\a\mu[\b}*\vt_{\g]}\,.
\ern
Therefore 
\begin{equation}\label{4.13}
\d C_{\a\b\g}*1=\d (d\vt_\a)\wedge *\vt_{\b\g}
+2\d\vt^\mu \wedge C_{\a\mu[\b}*\vt_{\g]}\,.
\end{equation}
After substitution of (\ref{4.12}--\ref{4.13}) into (\ref{4.11})
the variation of the Lagrangian  takes the form
\brn
\d L=
\frac 1{2} F^{\a\b\g}\Big(\d (d\vt_{\a})\wedge*\vt_{\b\g}+2\d\vt^\mu\wedge C_{\a\mu[\b}*\vt_{\g]}\Big)
+\d \vt^\mu \wedge (e_\mu\rfloor L)\,.
\ern
Extract the total derivatives  to obtain
 \brr\label{4.15}
\d L =\frac 1{2}
\d\vt_{\mu}\wedge\Big( d(*F^{\mu\b\g}\vt_{\b\g})+
2F^{\a\b\g}C_{\a\mu[\b}*\vt_{\g]}
+2 e_\mu\rfloor L\Big)%\nonumber\\&&
+\frac 1{2}d\Big(\d\vt_{\a}\wedge*F^{\a\b\g}\vt_{\b\g}\Big)\,.
\err
The variation relation (\ref{4.15})  plays a basic role in derivation of the field equation and of the conserved current. 
We rewrite it in a compact form by using the 2-forms (\ref{4.16}) and 
(\ref{4.16a}). 
The terms of the form $F\cdot C$  can be rewritten as 
\brn
F^{\a\b\g}C_{\a\mu[\b}*\vt_{\g]}=(F^{\a\b\g}-F^{\a\b\g})C_{\a\mu[\b}*\vt_{\g]}
%\nonumber\\
%&&\qquad 
= C_{\a\mu\b}*(e^\b\rfloor \F^\a)=-(e_\mu\rfloor \C_\a)\wedge *\F^\a\,.
\ern
Hence, (\ref{4.15}) takes the form 
 \brr\label{4.18}
\d L&=& \d\vt^\mu\wedge \Big(d(*\F_\mu)-(e_\mu\rfloor \C_\a)\wedge *\F^\a
+e_\mu\rfloor L\Big)+d(\d\vt^\mu\wedge \F_\mu)\,.
\err
Collect now the quadratic terms  into a differential 3-form 
 \begin{equation}\label{4.20}
\T_\m:=(e_\mu\rfloor \C_\a)\wedge *\F^\a - e_\mu\rfloor L\,.
 \end{equation}
Consequently,   the variational relation (\ref{4.15}) 
results in a compact form
\begin{equation}\label{4.22}
\d L= \d\vt^\mu\wedge \Big(d*\F_\mu-\T_\mu\Big)+
d\Big(\d\vt^\mu\wedge \F_\mu\Big)\,.
\end{equation}

%%%%%%%%%%%%%%%%%%%%%%%%%%%%%%%%%%%%%%%%%%%%%%%%%%%%
 \subsection{The coframe field equations}                 %%%%%%% 4
%%%%%%%%%%%%%%%%%%%%%%%%%%%%%%%%%%%%%%%%%%%%%%%%%%%%
We are ready now to write down the field equations.
Consider independent free variations of a  coframe field vanishing at infinity 
(or at the boundary of the manifold $\partial M$).  
The variational relation (\ref{4.22}) yields {\it the coframe field equation}   
\begin{equation}\label{4.24} 
d*\F^\mu=\T^\mu\,.
\end{equation}
Observe that the structure of coframe field equation is formally similar to the 
structure of the standard  electromagnetic field equation $d*F=J$. 
Namely, in both equations, the left hand sides  
are the exterior derivative of the dual field strength  while the right hand sides 
are odd 3-forms. 
Thus the 3-forms $\T^\mu$ serves as a source for the field strength  $\F^\mu$, 
as well as the 3-form of electromagnetic current $J$ is a source for the 
electromagnetic field $F$. 
There are, however, some important distinctions: 
(i) The coframe field current $\T_\mu$ is a vector-valued 3-form while 
the electromagnetic current $J$ is a scalar-valued 3-form. 
(ii) The field equation (\ref{4.24}) is nonlinear. 
(iii) The electromagnetic current $J$   depends on an exterior matter field, 
while the coframe current $\T^\mu$ is interior (depends on the coframe itself). 

The exterior derivation  of the both sides of 
field equation (\ref{4.24}) yields the conservation law
\begin{equation}\label{cons-I}
d\T^\mu=0.
\end{equation}
Note, that this equation obeys all the symmetries of the  coframe Lagrangian. 
It is diffeomorphism invariant and global $SO(1,3)$ covariant.  
Thus we obtain a conserved total 3-form (\ref{4.20}) which is 
constructed from the first order derivatives of the field variables (coframe). 
It is local and covariant.  
The 3-form $\T_\mu$ is our  candidate for the coframe energy-momentum current. 
%%%%%%%%%%%%%%%%%%%%%%%%%%%%%%%%%
\subsection{Conserved current and Noether charge}                       %%%%%%% 5
%%%%%%%%%%%%%%%%%%%%%%%%%%%%%%%%%
The current $\T_\mu$ is obtained 
directly, i.e., by separation of the terms in  the field equation.   
In order to identify the proper nature of this conserved 3-form  
we have to answer the question: 
{\bf{\it What symmetry this conserved current can be associated with?}} 

Return to the variational relation (\ref{4.22}). 
On shell, for the fields satisfying the field equations (\ref{4.24}), 
it takes the form
\begin{equation}\label{4.26}   
\d L=d (\d\vt^\a\wedge *\F_\a)\,.
\end{equation}
Consider the variations of the coframe field produced by the 
Lie derivative taken relative to a smooth vector field $X$, i.e., 
\begin{equation}\label{4.28a} 
\d\vt^\a=\L_X\vt^\a=d(X\rfloor\vt^\a)+X\rfloor d\vt^\a\,.
\end{equation}
The Lagrangian (\ref{4.10}) is a diffeomorphic invariant, 
hence its  variation is produced by the Lie  derivative taken relative 
to the same vector field $X$, i.e., 
\begin{equation}\label{4.28b} 
\d L=\L_X L=d(X\rfloor L)\,. 
\end{equation}
Thus the relation (\ref{4.26}) takes a 
form of a conservation law $d\Theta(X)$ for the Noether 3-form
\begin{equation}\label{4.29} 
\Theta(X)=\Big(d(X\rfloor\vt^\a)+X\rfloor \C^\a \Big)\wedge *\F_\a-X\rfloor L\,.
\end{equation}
This  quantity includes the derivatives of an   
arbitrary vector field $X$. 
Such a non-algebraic dependence  of the conserved current is an obstacle 
for definition of an energy-momentum tensor. 
This problem is solved 
merely by using the canonical form of the current. 
Let us  take $X=e_\a$. The first term of  (\ref{4.29})  vanishes  
identically.  Thus  
\begin{equation}\label{4.30} 
\Theta(e_\mu)=(e_\mu\rfloor \C^\a )\wedge *\F_\a-e_\mu\rfloor L\,.
\end{equation}
Observe that  the right hand side of the equation (\ref{4.30}) is 
exactly the same expression as
 the source term of the field equation (\ref{4.24}): 
\begin{equation}\label{4.30a} 
\Theta(e_\mu)=\T_\mu\,.
\end{equation}
Thus the conserved current $\T_\m$ defined in (\ref{4.20}) 
is associated with the diffeomorphism invariance of the Lagrangian. 
Consequently the vector-valued 3-form (\ref{4.20}) represents 
the {\it energy-momentum current of the coframe field}. 
%%%%%%%%%%%%%%%%%%%%%%%%%%%%%%%%%
%\subsection{Noether charge}                       %%%%%%% 5
%%%%%%%%%%%%%%%%%%%%%%%%%%%%%%%%%

Let us look for  an additional information incorporated 
in the conserved current (\ref{4.29}).
Extract the total derivative to obtain
 \begin{equation}\label{4.31aa} 
\Theta(X)=d\Big((X\rfloor\vt^\a)*\F_\a\Big)-(X\rfloor\vt^\a) (d*\F_\a-\T_\a)\,. 
\end{equation}
Thus, up to the field equation (\ref{4.24}), the current $\T(X)$ represents  
a total derivative of a certain 2-form 
$\Theta(X)=dQ(X)$\,. 
This result is a special case of a general proposition due 
to Wald \cite{Wald2}   for  a diffeomorphic invariant Lagrangians. 
The  2-form  
\begin{equation}\label{4.31} 
Q(X)=(X\rfloor \vt^\a) *\F_\a\,.
\end{equation}
can be   referred to as the {\it Noether charge for the coframe field}. 
Consider $X=e_\a$ and denote $Q_\a:=Q(e_\a)$. 
From (\ref{4.31}) we obtain that this canonical Noether charge
of the coframe field coincides with the dual of the conjugate strength 
\begin{equation}\label{4.32} 
Q_\a=Q(e_\a)=*\F_\a\,.
\end{equation} 
In this way, the 2-form $\F_\a$, which was used above only as a technical 
device for expressing the  equations in a compact form, obtained now 
a meaningful description. 
Note, that the Noether charge plays an important role in Wald's treatment 
of the black hole entropy \cite{Wald2}.

%%%%%%%%%%%%%%%%%%%%%%%%%%%%%%%%%%%%%%%%%%%%%%%%%%%%
\subsection{Energy-momentum tensor}                       %%%%%%% 5
%%%%%%%%%%%%%%%%%%%%%%%%%%%%%%%%%%%%%%%%%%%%%%%%%%%%
In this section we construct an expressions for the energy-momentum tensor 
for the coframe field.   
Let us first introduce the notion of the energy-momentum tensor 
via the differential-form formalism. 
We are looking for a second rank tensor field of a type $(0,2)$.
Such  a tensor can always be treated as a  bilinear map
$
T: \ {\mathcal X(M)}\times{\mathcal X(M)} \to {\mathcal F(M)},
$ 
where ${\mathcal F(M)}$ is the algebra of $C^\infty$-functions on $M$
while ${\mathcal X(M)}$ is the ${\mathcal F(M)}$-module of vector fields
on $M$. 
The unique way to construct a scalar from a 3-form and a vector is 
is to take the Hodge dual of the 3-form and to 
contract the result by the vector. 
Consequently, we define the energy-momentum tensor as 
\begin{equation}\label{2.27}  %tem1}
T(X,Y):=Y\rfloor *\T(X)\,.
\end{equation}
Observe that this quantity is a tensor if and only if the 3-form current
$\T$ depends  linearly (algebraic) on the vector field $X$.
Certainly, $T(X,Y)$ is not symmetric in general. 
The antisymmetric part of the energy-momentum tensor is known from 
the Poincar{\'e} gauge theory \cite{Hehl:1976kj} to represent the spinorial current of the 
field. 
The canonical form of the energy-momentum $T_{\a\b}:=T(e_\a,e_\b)$ tensor is 
\begin{equation}\label{2.28}    %tem3}
T_{\a\b}=e_\b\rfloor *\T_\a\,.
\end{equation}
Another useful form of this tensor can be obtained from (\ref{2.28}) 
by applying the rule (\ref{a0-7}) 
\begin{equation}\label{2.29}
T_{\a\b}=*(\T_\a\wedge \vt_\b)\,.
\end{equation}
The familiar procedure of rising the indices by the Lorentz metric $\eta^{\a\b}$ 
produces two tensors of a type $(1,1)$
\begin{equation}\label{2.30}
{T_\a}^\b=*(\T_\a\wedge \vt^\b),
\quad {\textrm {and}} \quad
{T^\a}_{\b}=*(\T^\a\wedge \vt_\b)\,,
\end{equation}
which are different, in general.  
By applying the rule (\ref{a0-7}) the first relation of (\ref{2.30}) 
is converted into
\begin{equation}\label{2.32}
\T_\a={T_\a}^\b*\vt_\b\,.
\end{equation}
Thus, the components of the energy-momentum tensor are regarded as  
the coefficients of the current $\T_\a$ in the dual basis
$*\vt^\a$ of the vector space $\Omega^3$ of odd 3-forms.

In order to show that  (\ref{2.32}) conforms with the intuitive notion 
of the energy-momentum tensor
let us restrict to a flat manifold and represent   the 3-form conservation law 
as a  tensorial expression. 
Take a closed  coframe $d\vt^\a=0$, thus $d*\vt_\b=0$. 
From (\ref{2.32}) we derive 
\brn
d\T_\a=d{T_\a}^\b\wedge *\vt_\b  =-{{T_\a}^\b}_{,\b}*1\,.
\ern
Hence, in this approximation,  the differential-form conservation law $d\T_\a=0$
is equivalent to the tensorial conservation law ${{T_\a}^\b}_{,\b}=0$.

Apply now the definition (\ref{2.28}) to the conserved current
(\ref{4.20}) for the coframe field.
The  energy-momentum tensor $T_{\mu\nu}=e_\nu\rfloor*\T_\mu$ 
is derived in the form 
\begin{equation}\label{4.39}
T_{\mu\nu}=e_\nu\rfloor*\Big((e_\mu\rfloor\C_\a)\wedge*\F^\a-
\frac 12 e_\mu\rfloor(\C_\a\wedge*\F^\a)\Big)\,.
\end{equation}
Using (\ref{a0-7}) we rewrite the first term in (\ref{4.39}) as 
\brn
&&e_\nu\rfloor*\Big((e_\mu\rfloor\C_\a)\wedge*\F^\a\Big)=
%-*\Big(\vt_n\wedge*^2\Big[(e_m\rfloor\C_a)\wedge*\F^a\Big]\Big)\\
%&&% \ \ \ =*\Big((e_m\rfloor\C_a)\wedge*^2(\vt_n\wedge*\F^a)\Big)=
-*\Big((e_\mu\rfloor\C_\a)\wedge*(e_\nu\rfloor\F^\a)\Big)\,.
\ern
As for the second term in (\ref{4.39}) it takes the form 
\brn
&&-\frac 12 e_\nu\rfloor*\Big(e_\mu\rfloor(\C_\a\wedge*\F^\a)\Big)=
%\frac 12*\Big(\vt_n\wedge *^2(e_m\rfloor(\C_a\wedge*\F^a)\Big)\\
%&&%=-\frac 12*(\vt_n\rfloor*\vt_m)*(\C_a\wedge*\F^a)=
\frac 12\eta_{\mu\nu}*(\C_\a\wedge*\F^\a)\,.
\ern
Consequently the energy-momentum tensor for the coframe field is  
\begin{equation}\label{4.40}
T_{\mu\nu}=-*\Big((e_\mu\rfloor\C_\a)\wedge*(e_\nu\rfloor\F^\a)\Big)+
\frac 12\eta_{\mu\nu}*(\C_\a\wedge*\F^\a)\,.
\end{equation}
Observe that this expression is formally similar to the known  
expression for the energy-momentum tensor of the Maxwell 
electromagnetic field: 
%Indeed, apply the definition (\ref{2.28}) to the conserved current
%(\ref{2.32b}) for the electromagnetic  field.
%The canonical form of the electromagnetic energy-momentum tensor is 
\begin{equation}\label{2.34a}       
{}^{(\tt em)}T_{\mu\nu}=-*\Big((e_\mu\rfloor F)\wedge *(e_\nu\rfloor F)\Big)+
\frac 12\eta_{\mu\nu}*(F\wedge *F)\,.
\end{equation}
The form (\ref{2.34a}) is no more than an expression of the 
electromagnetic energy-momentum tensor in arbitrary frame. 
In a specific coordinate chart $\{x^\mu\}$ it is enough to take the coordinate 
basis vectors $e_a=\partial_\a$ and consider 
$T_{\a\b}:={}^{(e)}T(\partial_\a,\partial_\b)$ 
 to obtain the familiar  expression 
 \begin{equation}\label{2.35}  %%Check the sign ????
{}^{(\tt em)}T_{\a\b}=
 -F_{\a\mu}{F_\b}^\mu+\frac 14 \eta_{\a\b}F_{\mu\nu}F^{\mu\nu}\,.
\end{equation}
The electromagnetic energy-momentum  tensor is obviously traceless.  
The same property holds also for the coframe field tensor.  In fact, the coframe energy-momentum tensor defined by 
(\ref{4.40}) is traceless for  all  models described by the Lagrangian (\ref{3.2}), i.e., for all values 
of the parameters $\rho_i$. 
Indeed, compute the trace ${T^\mu}_\mu=T_{\mu\nu}\eta^{\mu\nu}$ of (\ref{4.40}): 
\brn
{}^{(\tt cof)}{T^\mu}_\mu&=&-*\Big((e_\mu\rfloor\C_\a)\wedge*(e^\mu\rfloor\F^\a)\Big)+2*(\C_\a\wedge*\F^\a)\\
%&&=*\Big((e_\mu\rfloor\C_\a)\wedge*^2(\vt^\mu\wedge*\F^\a)\Big)+2*(\C_\a\wedge*\F^\a)\\
&=&-*\Big(\vt^\mu\wedge(e_\mu\rfloor\C_\a)\wedge*\F^\a\Big)+2*(\C_\a\wedge*\F^\a)=0\,.
\ern
%In the latter equality the relation  (\ref{A.9}) was used. 

It is well known that the traceless of the energy-momentum tensor is 
associated  with the scale invariance of the Lagrangian. 
The rigid ($\lambda$ is a constant) scale transformation
$x^i\to \lambda x^i$, is considered acting on a 
matter field as $\phi\to \lambda^d\phi$, where $d$ is the 
dimension of the field. 
The transformation does not act, however, on the components 
of the metric tensor and 
on the frame (coframe) components. 
It is convenient to shift the change on the metric and on the frame 
(coframe)  components, i.e., to consider 
\begin{equation}\label{4-42bx}
g_{ij}\to \lambda^2g_{ij}\,, \qquad
 {\vt^\a}_{i}\to \lambda{\vt^\a}_{i}\,, \qquad \textrm{and}
 {e_\a}^{i}\to \lambda^{-1}{\vt_\a}^{i}\,
 \end{equation}
with no change of coordinates. 
In the coordinate free formalism the difference between two approaches is neglected and the transformation is 
\begin{equation}\label{4-42b}
g\to\lambda^2 g\,,\qquad \vt^\a\to \lambda\vt^\a\,, \qquad \textrm{and} 
\qquad e_\a\to \lambda^{-1}e_\a\,.
\end{equation}
The transformation law of the coframe Lagrangian is simple to obtain from the 
component-wise form (\ref{4.5}). 
Under the transformation (\ref{4-42b}) the volume element changes as 
$*1\to \lambda^4 *1$. 
As for the $C$-coefficients, they transform due to (\ref{4.3}) as ${C^a}_{bc}\to\lambda^{-1}{C^a}_{bc} $. 
Consequently, by (\ref{3.3}),  the transformation law of the Lagrangian 4-form is $L\to\lambda^2 L$, 
which is the same as for the Hilbert-Einstein Lagrangian 
$L_{HE}=R\sqrt{-g}d^4x\to\lambda^2 L_{HE}$. 
After rescaling the Planck length  the scale invariance is reinstated. 
Hence, for the pure coframe field  model the energy-momentum tensor have to 
be traceless in accordance with the proposition above. 
%%%%%%%%%%%%%%%%%%%%%%%%%%%%%%%%%%%%%%%%%%%%%%%%%%%%
 \subsection{The field equation for a general system}           
%%%%%%% 5
%%%%%%%%%%%%%%%%%%%%%%%%%%%%%%%%%%%%%%%%%%%%%%%%%%%%
The coframe field equation have been derived for a pure coframe field. 
Consider now a general minimally coupled system of  
a coframe field $\vt^\a$ and a matter field $\psi$. 
The matter field can be  a differential form of an arbitrary degree 
 and can carry arbitrary number of exterior and interior indices. 
Take the total Lagrangian of the system to be of the form ($\ell=$ Planck
length)
\begin{equation}\label{4-41}
L=\frac 1{\ell^2}{}^{(\tt cof)}L(\vt^\a,d\vt^\a)+{}^{(\tt mat)}L(\vt^\a,\psi, d\psi)\,,
\end{equation}
where the coframe Lagrangian ${}^{(\tt cof)}L$, defined by (\ref{3.2}), 
is of dimension length square.   
The matter Lagrangian ${}^{(\tt mat)}L$ is dimensionless. 

The minimal coupling means here the absence of coframe derivatives 
in the matter Lagrangian.  
Take the variation of (\ref{4-41}) relative to the coframe field $\vt^\a$ 
to obtain 
\begin{equation}\label{4-42}
\d L=\frac 1{\ell^2}\d\vt^\a\wedge\Big(d*\F_\a-{}^{(\tt cof)}\T_\a-{\ell^2}{}^{(\tt mat)}\T_\a\Big)\,,
\end{equation}
where the 3-form of coframe current is defined by (\ref{4.24}). 
The 3-form of matter current  is defined via the variation 
derivative of the matter Lagrangian taken relative to the coframe field $\vt^\a$:
\begin{equation}\label{4-42aa}
{}^{(\tt mat)}\T_\a:=-\frac {\d}{\d\vt^\a}{}^{(\tt mat)}L\,.
\end{equation}
Introduce the total current of the system 
${}^{(\tt tot)}\T_\a={}^{(\tt cof)}\T_\a+{\ell^2}{}^{(\tt mat)}\T_\a$\,,
which is of dimension length (mass). 
Consequently, the field equation for the general system (\ref{4-41}) takes the 
form 
\begin{equation}\label{4-43}
d*\F_\a={}^{(\tt tot)}\T_\a\,.
\end{equation}
Using the energy-momentum tensor 
(\ref{2.32})  this equation can be rewritten in a tensorial form 
\begin{equation}\label{4-44}
e_\b\rfloor *d*\F_\a={}^{(\tt tot)}T_{\a\b}\,,
\end{equation}
or equivalently 
\begin{equation}\label{4-45}
\vt_\b\wedge d*\F_\a={}^{(\tt tot)}T_{\a\b}*1\,.
\end{equation}
The conservation law for the total current $d\T_\a=0$ is a straightforward 
consequence of the field equation (\ref{4-43}). 
The form (\ref{4-43}) of the field equation looks like 
 the Maxwell field equation for the 
electromagnetic field $d*F=J$. 
Observe, however, an important difference. 
The source term in the right hand side of the electromagnetic field equation 
depends only on external fields. 
In the absence of the external sources $J=0$, 
the electromagnetic strength $*F$ is a closed form. 
As a consequence, its cohomology class interpreted as a charge of the 
source. 
The electromagnetic field itself is uncharged.

As for the coframe field strength $\F^\a$ its source depends on the 
coframe and of its first order derivatives. 
Consequently, the 2-form $*\F^\a$ is not closed even in absence 
of the external sources. 
Hence the gravitational field is massive (charged) itself. 

On the other hand the tensorial form (\ref{4-44}) of the coframe  
field equation is similar to the 
Einstein field equation for the metric tensor
$
G_{\a\b}=8\pi {}^{(\tt mat)}T_{\a\b}\,.
$
Indeed, the left hand side in both equations are pure geometric 
quantities. 
Again, the source terms in the field equations  
are  different. 
The source of the Einstein gravity is the energy-momentum tensor  
only of the  matter fields. 
The conservation of this tensor is a consequence of the field equation. 
Thus even if  some meaningful conserved energy-momentum current for 
the metric field existed  
it would have been conserved regardless of  the matter field current. 
Consequently, any redistribution of the energy-momentum current between the 
matter 
and gravitational fields is  forbidden in the framework of the traditional 
Einstein gravity.  

As for the coframe field equation, the total energy-momentum current plays 
a role of the source of the field. 
Consequently the coframe field is completely ``self-interacted'' - the 
energy-momentum current of the coframe field produces an additional field. 
The conserved current of the coframe-matter system  is the total 
energy-momentum current, not only the matter current. 
Thus in the framework of general coframe  construction 
the redistribution of the energy-momentum current between  the matter field and the 
coframe field is possible, in principle. 
%---------------------
\subsection{Spherically symmetric solution}
%---------------------
 Let us look for a static spherically symmetric solution to the field equation (\ref{4.24}).
We will use the isotropic coordinates $\{x^\hi\,,\,\hi=1,2,3\}$ with the isotropic radius $\rho$. Denote
\begin{equation}
 s=\rho^2=\d_{\hi\hj}x^\hi x^\hj=x^2+y^2+z^2\,.
\end{equation}

 Recall that we identify the gravity variable with the coframe field defined
 up to an infinitesimal  Lorentz transformation.
 It is equivalent to the metric field.
So it is enough to look for a coframe solution  of
 a ``diagonal'' form  \cite{Itin:1999wi}
\begin{equation}\label{gr-anz}
\vt^0=f(s)\,dx^0\,,\qquad \vt^{\hi}=
g(s)\,dx^{\hi}\,.
\end{equation}%\labe{gr-anz}
Although this ansatz is not the most general one, it is enough because (\ref{gr-anz})  corresponds to a most general static spherical symmetric metric  
\begin{equation}\label{3-15}
ds^2=e^{2f(s)}dt^2-e^{2g(s)}(dx^2+dy^2+dz^2)\,.
\end{equation}
Substitution of (\ref{gr-anz}) into the field equation (\ref{4.24}) we  obtain an over-determined system of three second order ODE for two independent variables $f(s)$ and $g(s)$
\begin{equation}\label{5-2}
\left\{
\begin{array}{l}
\rho_1\Big(2f''s+3f'+2f'g's-2(g')^2s+(f')^2s\Big)+
2\rho_3\Big(2g''s+3g'+(g')^2s\Big)=0\\
\rho_1\Big(2g''+2f'g'-2(f')^2-2(g')^2\Big)+
2\rho_3\Big(f''+g''+(f')^2-2f'g'-(g')^2\Big)=0\\
\rho_1\Big(4g''s+4g'+4f'g's-2(f')^2s\Big)+
2\rho_3\Big(2f''s+2f'+2g''s+2g'+2(f')^2s\Big)=0.
\end{array}\right.
\end{equation}
This system has a solutions with the Newtonian behavior on infinity $f\sim 1-C/\rho$ only if the  parameter $\rho_1$ is equal to zero. In this case, the system (\ref{5-2}) has a unique solution 
\begin{equation}\label{6-5}
f=\ln\frac{1-\frac 1{c\rho}}{1+\frac 1{c\rho}},\qquad g=2\ln{\left(1+\frac 1{c\rho}\right)}\,.
\end{equation}
By taking the parameter of integration to be inversely proportional to the mass of the central body $c=\frac 2m$  we obtain the coframe field in the form
\begin{equation}\label{6-6}
\vt^0=\frac{1-\frac m{2\rho}}{1+\frac m{2\rho}}dt,\qquad \vt^i=\left(1+\frac m{2\rho}\right)^2dx^i, \qquad i=1,2,3\,.
\end{equation}
This coframe field yields the Schwarzschild metric in isotropic coordinates
\begin{equation}\label{6-7}
ds^2=\bigg(\frac{1-\frac m{2\rho}}{1+\frac m{2\rho}}\bigg)^2dt^2-\Big(1+\frac m{2\rho}\Big)^4(dx^2+dy^2+dz^2)\,.
\end{equation}
Note that the values of the parameters $\rho_2, \rho_3$ are not determined via the ``diagonal'' ansatz. Thus the Schwarzschild metric is a solution for a family of the coframe  field equations which defined by the parameters: 
\begin{equation}\label{6-7x}
\rho_1=0\,,\quad \rho_2,\rho_3 - {\rm arbitrary.}
\end{equation}
The ordinary GR is extracted from this  family  by
requiring of the {\it local} $SO(1,3)$ invariance, which is realized by 
an additional restriction of the parameters:
\begin{equation}\label{add1-2}
\rho_1=0\,,\quad 2\rho_2+ \rho_3=0 \,.
\end{equation}
%---------------------------
\subsection{Weak field approximation}
%---------------------------
Linear approximation of coframe models was usual applied  for study the deviation 
  from the standard GR, and for comparison with the 
 observation data, 
 see  \cite{Sezgin:1979zf}, \cite{Nitsch:1979qn},  
 \cite{Kuhfuss:1986rb}. We will use this approach to study the meaning of the condition 
 $\rho_1=0$, see \cite{Itin:2004ig}. Recall that this condition guarantees the existence of viable solutions. 
 
 To study the approximate solutions to (\ref{4.31}), we start with a
trivial exact solution, a {\it holonomic coframe}, for which 
\begin{equation}\label{la.1}
d{\vt}^a=0\,.
\end{equation}
Consequently, $\F^a=\C^a=0$, so both sides of Eq. (\ref{4.31})  vanish.
By Poincar{\'e}'s lemma, the solution of (\ref{la.1}) can be locally expressed as
${\vt}^a=d\tilde{x}^a(x)$, where  $\tilde{x}^a(x)$ is a set of four
smooth functions defined in a some neighborhood $U$ of a point $x\in\M$.
The functions $\tilde{x}^a(x)$,
being treated as  the components of a coordinate map
$\tilde{x}^a:U \to{\mathbb R}^4$, generate a local coordinate system on $U$.
The metric tensor  reduces, in this
coordinate chart,  to the  flat Minkowski metric 
$g=\eta_{ab}d\tilde{x}^a\otimes d\tilde{x}^b$.
Thus the holonomic coframe plays, in the coframe background, the same role as the
Mankowski metric in the (pseudo-)Riemannian geometry.
Moreover, a manifold endowed with a (pseudo-)orthonormal holonomic
coframe is  flat. 
The weak perturbations of the basic solution $\vt^a=dx^a$ are  
\begin{equation}\label{la.2}
\vt^a=dx^a+h^a=(\d^a_b+{h^a}_b)\,dx^b\,.
\end{equation}
The indices in ${h^a}_{b}$ can be lowered and raised 
by the Mankowski metric 
\begin{equation}\label{la.7}
h_{ab}:=\eta_{am}{h^m}_{b}\,, \qquad h^{ab}:=\eta^{bm}{h^a}_{m}\,.
\end{equation}
The first operation is exact (covariant to all orders of approximations), while 
the second  is covariant only  to the first order, when $g^{ab}\approx \eta^{ab}$. 
The symmetric and the antisymmetric combinations of the  perturbations
\begin{equation}\label{la.9}
\theta_{ab}:=h_{(ab)}=\frac 12 (h_{ab}+h_{ba}),\qquad {\text{and}}\qquad
w_{ab}:=h_{[ab]}=\frac 12 (h_{ab}-h_{ba})\,.
\end{equation}
as well as the  trace
%\begin{equation}\label{la.11}
$\theta:={h^m}_m={\theta^m}_m$
%\end{equation}
are covariant to the first order. 
 The components of the metric tensor, in the linear approximation, involve only the
symmetric combination of the coframe perturbations
\begin{equation}\label{la.12}
g_{ab}=\eta_{ab}+2\theta_{ab}\,.
\end{equation}
When the decomposition 
\begin{equation}\label{la.12x}
h_{ab}=\theta_{ab}+w_{ab}\,
\end{equation}
is applied, 
the  field strength is splitted  to a sum of two
independent strengths --- one defined by  the symmetric field $\theta_{ab}$ 
and the second one defined by the antisymmetric field $w_{ab}$
\begin{equation}\label{fs.5}
\F_a(\theta_{mn},w_{mn})={}^{(\tt sym)}\F_a(\theta_{mn})+{}^{(\tt ant)}\F_a(w_{mn})\,,
\end{equation}
where
\begin{equation}\label{fs.5a}
{}^{(\tt sym)}\F_a=-\left[(\rho_1+\rho_3)\theta_{a[b,c]}+
\rho_3\eta_{a[b}{\theta_{c]m}}^{,m}-\rho_3\eta_{a[b}\theta_{,c]}\right]\vt^b\wedge \vt^c\,,
\end{equation}
and
\begin{equation}\label{fs.5b}
{}^{(\tt ant)}\F_a=-\left[(\rho_1+\rho_3)w_{a[b,c]}+3\rho_2w_{[ab,c]}-
\rho_3\eta_{a[b}{w_{c]m}}^{,m}\right]\vt^b\wedge \vt^c\,.
\end{equation}
Hence, for arbitrary values of the parameters $\rho_i$, the field
strengths of the fields $\theta_{ab}$ and $w_{ab}$ are independent. 

The linearized field equation takes the form 
%\br \label{sfe.3x}%CHECK!
%&&\rho_3\Big(2{h_{a[b,c]}}^{,c}-\eta_{ac}{h_{mb}}^{,m,c}+\eta_{ab}{h_{mc}}^{,m,c}+
%{h_{,a,b}}-\eta_{ab}\square h\Big)+6\rho_2{w_{[ab,c]}}^{,c}=0\,.
%\er
%Let us separate  the derivatives of $\theta_{ab}$ and the derivatives of $w_{ab}$
\begin{eqnarray} \label{sfe.3}
&& (\rho_1+\rho_3)\big(\square\,\theta_{ab}-{\theta_{am,b}}^{,m}\big)+
\rho_3\big(-\eta_{ab}\square\,\theta
  -{{\theta_{mb}}^{,m}}_{,a}+\theta_{,a,b}+\eta_{ab}{\theta_{mn}}^{,m,n}\big)+\nonumber\\
&& \qquad  (\rho_1+2\rho_2+\rho_3) \big(\square \,w_{ab}-{w_{am,b}}^{,m}\big)+
(2\rho_2+\rho_3){w_{bm,a}}^{,m}=0\,.
\end{eqnarray}
%Recall, that, in the linear approximation, the coframe field is reduced
%covariantly (to the first order) to a superposition of two independent
%fields.
%We are looking now for which values of the parameters the
%corresponding reduction appears
%in the field equation (\ref{sfe.3}).

 \vspace{0.3cm}

 \noindent {\bf Proposition 1:}
{\it 
For the case  $\rho_1=0$,  the linearized coframe field equation (\ref{sfe.3}), splits, 
in arbitrary coordinates,   into  two independent systems
$${}^{(\tt sym)}{\cal{E}}_{(ab)}(\theta_{mn})=\square\, \theta_{ab}=0\,,
\qquad {\textrm {and}} \qquad
{}^{(\tt ant)}{\cal{E}}_{[ab]}(w_{mn})=\square \,w_{ab}=0\,.$$
If  $\rho_1\ne 0$, Eq.(\ref{sfe.3}) does not split in
any coordinate system.}

Consequently, for $\rho_1=0$ and for generic values of the parameters $\rho_2,\rho_3$,
the field equation of the coframe field is splitted 
to two independent field equations for two independent field variables. 
This splitting emerges also for the Lagrangian and the energy-momentum current. 

 \vspace{0.3cm}

 \noindent {\bf Proposition 2:}
{\it For $\rho_1=0$, the Lagrangian of the coframe field
is reduced, up to a total derivative term, to the sum of two independent Lagrangians
\begin{equation}\label{rl.2}
\L(\theta_{ab},w_{ab})={}^{\tt (sym)}\L(\theta_{ab})+{}^{\tt (ant)}\L(w_{ab})\,.
\end{equation}
Moreover, the coframe energy-momentum current  is reduced, on shell, in the first order approximation, as
\begin{equation}\label{em.1}
\T_a(\theta_{mn},w_{mn})= {}^{\tt (sym)}\T_a(\theta_{mn})+{}^{\tt (ant)}\T_a(w_{mn})\,,
\end{equation}
up to a total derivative.}

The result of our analysis is as following: In the linear approximation the field variable is splitted to a sum of two independent fields. These fields do not interact only in the case $\rho_1=0$. Remarkable that this condition coincides with the viable condition (\ref{6-7x}), which is necessary for Schwarzschild metric. 

%--------------------------------------------
 \section{Coframe geometry}
%--------------------------------------------
The coframe gravity represented above is not related to a certain specific geometric structure. In this section we are looking for a geometry that can be constructed from the coframe field.  
It is well known that, on a Riemannian manifold there exists a unique linear connection of Levi-Civita \cite{kob1}. Already this statement  indicates that when we want to deal with some other connection, for instance with the flat one, we have to use some other  non-Riemannian geometric structure. In this section, we  define a geometry based on a coframe field. It is  instead of the of the standard Riemannian geometry based on a metric tensor field.
%--------------------------------------------
\subsection{Coframe manifold. Definitions and notations}
%--------------------------------------------
Our construction will repeat the main properties of the Riemannian structure. Let us start with the basic definitions.

 \vspace{0.3cm}

%--------------------------------------------
\noindent{\bf{ Differential manifold.}}
 Let $M$ be a smooth $n+1$ dimensional differentiable manifold, which is locally (in an open set $U\subset M$)  parametrized by a coordinate chart $\{x^i;\,i=0, 1, \ldots n\}$.
The set of $n+1$ differentials $dx^i$ provides a coordinate basis for the module of the differential forms on $U$. Similarly, the set of $n+1$ vector fields $\partial_i=\partial/\partial x^i$ forms the coordinate basis for the module of the vector fields on $U$. Arbitrary smooth transformations  of the coordinates $x^i\to y^i(x^j)$ are admissible. Under these transformations, the elements of the coordinate bases transform by the tensorial law
\begin{equation}\label{Dif-0}
dx^i\to dy^i=\frac{\partial y^i}{\partial x^j}\, dx^j\,, \qquad \frac{\partial}{\partial x^i}\to \frac{\partial}{\partial y^i}=\frac{\partial x^j}{\partial y^i}\,\frac{\partial}{\partial x^j}\,.
\end{equation}
The Jacobian matrix ${\partial y^i}/{\partial x^j}$ is assumed to be smooth and invertible.
The coordinate bases $dx^i$ and $\partial_i=\partial/\partial x^i$ are referred to as {\it holonomic bases}. They satisfy the relations $d(dx^i)=0$ and $[\partial_i,\partial_j]=0$.

For a compact representation of geometric quantities, it is useful to have an alternative description via {\it nonholonomic bases}.
Denote by $\theta^a$ a generic nonholonomic basis of the module of the 1-forms on
$U$. Its dual $f_a$ is a basis of of the module of the vector fields  on $U$.
In general, $d\theta^a\ne 0$ and $[f_a,f_b]\ne 0$.
Relative to the coordinate bases, the elements of the nonholonomic bases are locally expressed as
\begin{equation}\label{Dif-1}
\theta^a=\theta^a{}_i\,dx^i\,,\qquad f_a=f_a{}^i\,\partial_i\,.
\end{equation}
Here the matrices $\theta^a{}_i$ and $f_a{}^i$ are the inverse to each-other, i.e.,
\begin{equation}\label{Dif-1x}
\theta^a{}_i\,f_a{}^j=\delta_i^j\,,\qquad \theta^a{}_i\,f_b{}^i=\delta_b^a\,.
\end{equation}
 Arbitrary smooth pointwise transformations of the nonholonomic bases
 \begin{equation}\label{Dif-2}
\theta^a\to A^a{}_b(x)\theta^b\,,\qquad f_a\to A_a{}^b(x)f_b\,.
\end{equation}
are admissible. Here $A_a{}^b$  denotes, as usual, the matrix inverse to $A^a{}_b$.

 Although the  basis indices change in the same range $a,b,\cdots=\{0, 1, \ldots, n\}$ they are distinguished from the coordinate indices $i,j,\cdots$. In particular, the contraction of the indices in the quantities $\theta^a{}_i$ or $f_a{}^i$ is forbidden since the result of such an action is not a scalar.
 The  base transformations (\ref{Dif-2}) are similar to the coordinate transformations (\ref{Dif-0}).
Note that the basis $\theta^a$ can be changed to an arbitrary other basis, for instance to the coordinate one. Indeed, the formulas (\ref{Dif-1}) can be treated as certain transformations of the bases.
 Consequently,  $\theta^a$ cannot be given any intrinsic geometrical sense. In particular, it cannot be used as  a model of a physical field.

 \vspace{0.3cm}

\noindent{\bf{Coframe field.}} Let the manifold $M$ be endowed with a smooth nondegenerate coframe field $\vt^\a$. It comes together with its dual --- the frame field $e_\a$.
In an arbitrary chart of  local coordinates $\{x^i\}$, these
fields are expressed as
 \begin{equation}\label{Cofrfield1}
 \vt^\a={\vt^\a}_i dx^i\,,\qquad e_\a={e_\a}^{i}\partial_i\,,
\end{equation}
 i.e., by two nondegenerate matrices ${\vt^\a}_i$ and ${e_\a}^i$ which are the inverse to each-other. In other words, we are considering a set of $n^2$ independent smooth functions on $M$.
Also the coframe indices change in the same  range $\a,\b,\ldots=0,\ldots, n$ as the coordinate indices $i,j,\ldots$ and  the basis indices $a, b, \ldots$. They all however have to be strictly distinguished. In particular, the indices in ${\vt^\a}_i$ or ${e_\a}^{i}$
 cannot be  contracted.
 %---------------------

 \vspace{0.3cm}

\noindent {\bf{Coframe transformation.}}
 %---------------------
 For  most physical models  based on the coframe field, this field is defined only up to global transformations.
It is natural to consider a wider class of coframe fields related by  local pointwise transformations
\begin{equation}\label{Transprop1}
 \vt^\a\to L^\a{}_{\b}(x)\vt^\b\,,\qquad e_\a\to L_\a{}^{\b}(x)e_\b\,.
\end{equation}
Here $L^\a{}_{\b}(x)$ and $L_\a{}^{\b}(x)$ are  inverse to each-other at arbitrary point $x$. Denote the group of matrices $L^\a{}_{\b}(x)$ by $G$.
Note two specially important cases: (i) $G$ is a group of global transformations with a constant matrix $L^\a{}_{\b}$; (ii)  $G$ is a group of arbitrary local transformations  such that the entries of $L^\a{}_{\b}$ are arbitrary functions of a point. In the latter case, the difference between the coframe field $\vt^\a$ and the reference basis $\theta^a$ is completely removed and the coframe structure is trivialized.

Consequently we involve an additional element of the coframe structure --- {\it the coframe transformations group}
\begin{equation}\label{Transprop2}
 G=\Big{\{}L^\a{}_{\b}(x)\in GL(n+1, \mathbb R);\, {\text{for\, every}} \, x\in M\Big{\}}\,.
\end{equation}
On this stage, we only require the matrices $L^\a{}_{\b}(x)$ to be invertible at an arbitrary  point $x\in M$. The successive specializations of the coframe transformation  matrix  will be involved  in sequel.

%--------------------------------

 \vspace{0.3cm}

\noindent{\bf{Coframe field volume element.}}
%--------------------------------
 We assume the coframe field to be non-degenerate at  an arbitrary point $x\in M$.
Consequently,  a special $n+1$-form, {\it the coframe field volume element}, is defined and nonzero.
Define
\begin{equation}\label{volel1}
{\rm {vol}}(\vt^\a)=\frac 1{n!}\,\varepsilon_{\a_o\cdots
\a_{n}}\vt^{\a_o} \wedge\cdots\wedge\vt^{\a_{n}}\,,
\end{equation}
where
$\varepsilon_{\a_o\cdots \a_{n}}$ is the  Levi-Civita
permutation symbol normalized by $\varepsilon_{01\cdots n}=1$.
Treating  the coframe volume element as one of the basic elements of the coframe geometric structure, we apply the following invariance condition.

\begin{itemize}
\item[]{\underline{\it Volume element invariance postulate:}}
 Volume element ${\rm {vol}}(\vt^\a)$ is assumed to be invariant under pointwise transformations of the coframe field
\begin{equation}\label{volel2}
{\rm {vol}}\left(\vt^\a\right)={\rm {vol}}\left(L^{\a}{}_\b\vt^\b\right)\,.
\end{equation}
%\end{itemize}
This condition is satisfied by  matrices  with unit determinant. Consequently, the coframe transformation group (\ref{Transprop2}) is restricted to
\begin{equation}\label{volel2x}
G=\Big{\{}L^\a{}_{\b}(x)\in SL(n+1, \mathbb R);\, {\text{for\, every}} \, x\in M\Big{\}}\,.
\end{equation}
\end{itemize}

 \vspace{0.3cm}

\noindent{\bf{Metric tensor.}} For a meaningful physical field model, it is necessary to have a
 metric structure on $M$. Moreover, the metric tensor
 has to be  of the Lorentzian signature.
 In a coordinate basis  and
 in an arbitrary reference basis, a generic metric tensor  is written correspondingly as
 \begin{equation}\label{Metrtensor1}
 g=g_{ij}dx^i\otimes dx^j\,,\qquad
 g=g_{ab}\theta^a\otimes\theta^b\,,
 \end{equation}
 where the components $g_{ij}$ and $g_{ab}$ are smooth functions
 of a point $x\in M$.

 On a coframe manifold, a metric tensor is not an independent quantity. Instead, we are looking for a metric explicitly constructed from
 a given  coframe field, $g=g(\vt^\a)$.
 We assume the metric tensor to be quadratic
 in the coframe field components and independent of its derivatives. Moreover,
 it should be of the Lorentzian type, i.e., should be reducible at a point to  the Lorentzian   metric $\eta_{\a\b}= {\rm{diag}}(-1,1,\cdots,1)$.
These requirements are justified by an almost flat approximation: for an almost holonomic coframe, $\vt^\a{}_i\approx \d^\a_i$, we have to reach the flat Lorentzian metric.
With these restrictions, we come to a definition of the {\it coframe field metric tensor} 
\begin{equation}\label{Metrtensor2}
 g=\eta_{\a\b}\vt^\a\otimes \vt^\b\,, \qquad  g_{ij}=\eta_{\a\b}\vt^\a{}_i\vt^\b{}_j\,.
 \end{equation}
 Note that the equations (\ref{Metrtensor2}) often
 appear  as a definition of a (non unique) orthonormal basis of
 reference for a given metric.
 Another interpretation treats (\ref{Metrtensor2}) as an expression
 of a given metric in  a special orthonormal basis of reference, as in (\ref{Metrtensor1}).
 In our approach, (\ref{Metrtensor2}) has a principle different meaning.
 It is  a definition of  the metric tensor field
 via the  coframe field.
 Certainly  the form of the metric $\eta_{\a\b}$ in the tangential
 vector space $T_xM$ is an additional axiom of our
 construction.
With an aim to define an invariant coframe geometric structure we require:
\begin{itemize}
\item[]{\underline{\it Metric tensor invariance postulate:}}
 Metric tensor  is assumed to be invariant under pointwise transformations of the coframe field, i.e.,
\begin{equation}\label{Metrtensor3}
g\left(\vt^\a\right)=g\left(L^\a{}_\b\vt^\b\right)\,.
\end{equation}
%\end{itemize}
This condition is satisfied by  pseudo-orthonormal matrices,
\begin{equation}\label{Metrtensor3x}
\eta_{\mu\nu}L^\mu{}_\a L^\nu_\b=\eta_{\a\b}.
\end{equation}
 Consequently, the  invariance of the  coframe metric restricts  the coframe transformation group to
\begin{equation}\label{Metrtensor4}
G=\Big{\{}L^\a{}_{\b}(x)\in O(1, n, \mathbb R);\, {\text{for\, every}} \, x\in M\Big{\}}\,.
\end{equation}
In order to have  simultaneously a metric and a volume element structures both constructed from the coframe field,  we have to assume a successive  restriction of the coframe transformation group:
\begin{equation}\label{Metrtensor5}
G=\Big{\{}L^\a{}_{\b}(x)\in SO(1, n, \mathbb R);\, {\text{for\, every}} \, x\in M\Big{\}}\,.
\end{equation}
\end{itemize}
 %--------------------------------

 \vspace{0.3cm}

\noindent{\bf{Topological restrictions.}} A global smooth coframe field may be defined only on a
parallelizable manifold, i.e., on a topological manifold of a zero
second Whitney class.  This topological restriction is equivalent
to existence of a spinorial structure on $M$.
 In this chapter, we restrict ourselves to a local consideration, thus
the global definiteness problems  will be neglected. Moreover, we
assume
 the coframe field to be  smooth and nonsingular only in a
"weak" sense. Namely, the components $\vt^\a{}_i$ and $e_\a{}^i$ are
 required to be differentiable and
linearly independent at almost all points of $M$, i.e., except
of a zero measure set.  So, in general, the coframe field can
degenerate at singular points, on singular lines (strings), or
even on singular submanifolds ($p$-branes).  This assumption
leaves a room for the standard singular solutions of the physics field equations such as
the Coulomb field, the Schwarzschild metric, the Kerr metric etc..

{\subsection{Coframe connections}}
%-------------------------------------------
From the geometrical point of view, a differential manifold endowed with a coframe field is  a rather poor  structure.
In particular,  we can not determine if two vectors attached at distance points
 are parallel to each-other or not.
 In order to have a meaningful geometry and, consequently, a
 meaningful geometrical field model for gravity, we have to consider a reacher
 structure.
 In this section we define a coframe manifold with a linear coframe
 connection. The connection 1-form $\GG_a{}^b$ will not be an independent
 variable, as in the Cartan geometry or in MAG \cite{Hehl:1994ue}. Alternatively in our construction the connection  will be     explicitly constructed
 from the  coframe field and its first order derivatives.
Thus we are dealing with a category of  {\it coframe manifolds with a linear coframe
 connection:}
 \begin{equation}\label{Cofcon_1}
 \Big{\{}M\,, \vt^\a,G\,,\G_a{}^b(\vt^\a)\Big{\}}\,.
 \end{equation}
We  start with a coframe manifold without an addition metric structure.
Metric contributions to the connection will be considered in sequel.

 %-----------------------

 \vspace{0.3cm}

\noindent{\bf{Affine connection.}} Recall the main properties of a generic linear affine connection on an $(n+1)$ dimensional differential manifold.
 Relative to a local coordinate chart
 $x^i$, a connection is represented by a set of $(n+1)^3$ independent
 functions ${\G^k{}_{ij}}(x)$ --- {\it the coefficients of the connection.}
   The only condition these functions have to satisfy is to
   transform, under a change of coordinates $x^i\mapsto y^i(x)$,
   by an inhomogeneous linear rule:
\begin{equation}\label{cc1}
\G^i{}_{jk}\mapsto
\left(\G^l{}_{mn}\frac{\partial y^m}{\partial x^j}\frac{\partial y^n}
{\partial x^k}+\frac{\partial^2 y^l}{\partial x^j\partial x^k}\right)
\frac{\partial x^i}{\partial y^l}\,.
 \end{equation}
 When  an arbitrary reference basis  $\{\theta^a\,,f_b\}$ is involved,
 the coefficients of the connection
 are arranged in a $GL(n,\mathbb R)$-valued {\it connection 1-form}, which is defined as
  \cite{kob1}
  \begin{equation}\label{cc2}
 \GG_a{}^b={f_a}^k\left({\theta^b}_i\G^i{}_{jk}{}-{\theta^b}_{k,j}\right)dx^j\,.
 \end{equation}
In  a holonomic coordinate basis, we can simply use the identities
$\theta^a{}_i=\delta^a_i$ and $f_a{}^i=\delta_a^i$.
Consequently, in a coordinate basis, the derivative term is canceled out and  (\ref{cc2}) reads 
 \begin{equation}\label{cc4}
 \GG_j{}^i=\G^i{}_{jk}{}\,dx^k\,.
 \end{equation}
 Due to (\ref{cc1}), this quantity transforms under the coordinate transformations as
 \begin{equation}\label{cc4x}
 \GG_j{}^i\to\left[\G_m{}^l\frac{\partial y^m}{\partial x^j}+
d\left(\frac{\partial y^l}{\partial x^j}\right)\right]\frac{\partial x^i}{\partial y^l}\,.
 \end{equation}
Alternatively, the connection 1-form   (\ref{cc2}) is  invariant under smooth transformations of
coordinates. The inhomogeneous linear behavior is shifted here to
the transformations of $\GG_a{}^b$ under a linear local map of
the reference basis $(\theta^a,f_a)$ given in (\ref{Transprop1}):
 \begin{equation}\label{cc3}
 \GG_a{}^b\mapsto
 \left({\G_c{}^d}  A_a{}^c+dA_a{}^d\right)A^b{}_d\,.
 \end{equation}
On a manifold with a given  coframe field $\vt^\a$, the connection
1-form (\ref{cc2}), can  also  be  referred to this  field. We
denote this quantity by ${\GG_\a}^\b$. It is defined similarly to
(\ref{cc2}):
 \begin{equation}\label{cc5}
 \GG_\a{}^\b=\left({\vt^\b}_i\G^i{}_{jk}-{\vt^\b}_{k,j}\right)
 {e_\a}^k\,dx^j\,.
 \end{equation}
 This quantity can be treated as an expression of a generic connection (\ref{cc2}) in a special basis. Note an essential difference between two very similar equations  (\ref{cc2}) and (\ref{cc5}). In (\ref{cc2}), we must be able to apply arbitrary pointwise linear transformations of the basis. The coefficients of the connection $\G^i{}_{jk}$ are independent on the basis $(\theta^a,f_a)$ used in  (\ref{cc2}). On the other hand, in
 (\ref{cc5}), we permit  only the transformations of the coframe field $\vt^\a$ that are  restricted by some invariance requirements. Moreover, we will require the connection $\G^i{}_{jk}$ to be constructed explicitly from the derivatives of the coframe field itself.

%--------------------------------------

 \vspace{0.3cm}

\noindent{\bf{Linear coframe connections. }}
 %--------------------------------------
We  restrict ourselves  to the quasi-linear $\G^i{}_{jk}(\vt^\a)$,
 i.e., we consider a connection   constructed as a linear
 combination of the
 first order derivatives of the coframe field. The coefficients in
 this linear expression  may depend on the frame/coframe components.
In other words, we are looking for a coframe analog of an ordinary Levi-Civita connection.

Let us assist ourselves  with a similar construction from the
 Riemannian geometry. So let us look now for a most general connection that can be constructed from the metric tensor components.
 Consider a general linear combination of the first order
 derivatives of the metric tensor:
 \begin{equation}\label{Lincofcon_1}
 g^{mi}(\a_1 g_{mj,k}+\a_2g_{mk,j}+\a_3g_{jk,m})\,.
 \end{equation}
 Although this expression has the same  index content as $\G^i{}_{jk}$, it is a connection only for some special values of the parameters $\a_1,\a_2.\a_3$.
 Indeed, any two  connections are differ by a tensor. Thus  an
 arbitrary
 connection can be expressed as a certain special connection plus a tensor
\begin{equation}\label{Lincofcon_2}
 \G^i{}_{jk}=\LG^i{}_{jk}+K^i{}_{jk}\,.
 \end{equation}
Use for  $\LG^i{}_{jk}$ the Levi-Civita connection
 \begin{equation}\label{Lincofcon_3}
 \LG^i{}_{jk}=\frac 12 g^{im}(g_{mj,k}+g_{mk,j}-g_{jk,m})\,.
 \end{equation}
 However in Riemannian geometry,
 does not exist a tensor constructed from the first order derivatives of the metric.
 Therefore $K^i{}_{jk}=0$, thus the Levi-Civita connection is a unique  connection that can be constructed from the first order derivatives of the metric tensor.
 It is evidently symmetric and metric compatible.

 In an analogy to this construction, we will look for a most general
 coframe connection of the form
\begin{equation}\label{Lincofcon_4}
 \G^i{}_{jk}(\vt^\a)=\oG^i{}_{jk}(\vt^\a)+K^i{}_{jk}(\vt^\a)\,.
 \end{equation}
Here $\oG^i{}_{jk}$ is a certain special  connection, while $K^i{}_{jk}$ is a tensor.
 To start with, we need a certain
 analog of the Levi-Civita connection, i.e., a special connection
 constructed from the coframe field.

%--------------------------------------

 \vspace{0.3cm}

\noindent{\bf{The flat Weitzenb\"{o}ck connection. }}
 %--------------------------------------
 On a bare differentiable manifold $M$, without any additional structure,
 the notion of parallelism of two vectors attached to
 distance points  depends on a curve joint
 the points.
 Oppositely, on a coframe manifold $\{M\,,\vt^\a\}$,
 a certain type of the  parallelism of distance vectors may be defined in
 an absolute (curve independent) sense \cite{Thomas}.
 Namely, two vectors $u(x_1)$ and $v(x_2)$ may be declared parallel to each other, if,
 being referred to the local elements of the coframe field $u(x_1)=u_\a(x_1)\vt^\a(x_1)$ and $v(x_2)=v_\a(x_2)\vt^\a(x_2)$,
 they have the proportional components $u_\a(x_1)=C v_\a(x_2)$.
 This definition is independent on the coordinates used on the manifold
 and on the nonholonomic frame of reference. It do depends on the coframe field.
 Since, by local transformations, the coframes at distance points
 change differently, only rigid   linear coframe
 transformations  preserve  such type of a  parallelism.

 This geometric picture may be reformulated
 in term of a special connection.
 The elements of the coframe field  attached to distinct points
 have to be  assumed parallel to each other.
 It means that a special connection $\oG^i{}_{jk}$ exists such that the
 corresponding covariant  derivative  of the
 coframe  field components  is zero:
 \begin{equation}\label{Weitz1}
 {\vt^\a}_{j;k}={\vt^\a}_{j,k}-\oG^i{}_{jk}{\vt^\a}_i=0\,.
 \end{equation}
 Multiplying by ${e_\a}^i$, we have an explicit expression
  \begin{equation}\label{Weitz2}
 \oG^i{}_{jk}={e_\a}^i{\vt^\a}_{k,j}\,.
 \end{equation}
 Under a smooth transform of coordinates, this expression
 is transformed in accordance with the inhomogeneous
 linear rule (\ref{cc1}). Consequently,
 (\ref{Weitz2}) indeed gives the coefficients of a special connection
 which is referred to as  the {\it Weitzenb\"{o}ck flat connection}.
This connection is unique for a
 class of coframes related by rigid linear transformations.

 In an arbitrary nonholonomic reference basis  $(\theta^a,f_a)$,
 we have correspondingly a
 unique Weitzenb\"{o}ck's connection 1-form which is constructed by
 (\ref{cc2}) from  (\ref{Weitz2})
  \begin{equation}\label{Weitz3}
\GG_a{}^b={f_a}^k\left({\theta^b}_i\oG^i{}_{jk}{}-{\theta^b}_{k,j}\right)dx^j\,.
 \end{equation}
Substituting the coframe field $\vt^\a$ instead of the nonholonomic basis $\theta^a$  we have
 \begin{equation}\label{Weitz3x}
 \oGG_\a{}^\b=\left(-{\vt^\b}_{i,j}+{\vt^\b}_k{e_\a}^k{\vt^\a}_{i,j}
 \right){e_\a}^i\,dx^j =0\,.
 \end{equation}
 Thus the Weitzenb\"{o}ck connection 1-form is zero,
 when it is  referred to the coframe field  $(\vt^\a, e_\a)$
 itself.
 Certainly, this property  is only a basis related fact.
 It yields, however, vanishing of  the
 curvature of the Weitzenb\"{o}ck connection,
 which is a basis independent property.
 %-------------------------------------------------------

%-------------------------------------------------------

 \vspace{0.3cm}

\noindent{\bf{General coframe connections. }}
%-------------------------------------------------------
 Recall that we are looking for a general coframe connection
 constructed from  the  first order derivatives of the coframe
 field components. In the Riemannian
 geometry, the analogous construction yields an unique
 connection of Levi-Civita. In the
 coframe geometry, however, the situation is different.

 \vspace{0.3cm}

 \noindent {\bf Proposition 3:}
{\it The general linear connection constructed from the
first order derivatives of the coframe field is given by a
3-parametric family:
  \begin{equation}\label{GCC1}
 \G^i{}_{jk}=\oG^i{}_{jk}+\a_1C^i{}_{jk}+\a_2C_j\d^i_k+
 \a_3C_k\d^i_j\,.
 \end{equation}
 }
 \noindent {\bf Proof:}
 The difference of two connections is a tensor  of a type $(1,2)$,
 so an arbitrary connection can be expressed as
 the Weitzenb\"{o}ck connection plus a tensor
 \begin{equation}\label{GCC2}
 \G^i{}_{jk}=\oG^i{}_{jk}+K^i{}_{jk}\,.
 \end{equation}
 Since $\oG^i{}_{jk}$ is already a linear combination of the first order
 derivatives,
 the additional tensor  also has to be of the same form.
 Observe that $K^i{}_{jk}$ involves only coordinate indices, while
 the partial derivatives ${\vt^\a}_{j,i}$ have a coframe index $\a$.
 This coframe index has to be suppressed.
 Hence the first order derivatives of the
 coframe components may appear in $K^i{}_{jk}$ only by
 the expressions ${e_\a}^i{\vt^\a}_{j,k}$.
 Notice that this quantity coincides with
 the coefficients  of Weitzenb\"{o}ck's connection (\ref{Weitz2}), which is not a tensor.
 Since the matrix of the frame field components $e_\a{}^j{}$ is the inverse of $\vt^\a{}_i$,
 the derivatives of the frame field $e_\a{}^j{}_{,k}$
 are  linear combinations of ${\vt^\a}_{j,k}$.
 Thus we do not need to involve  additional derivatives of the
 frame field into  $K^i{}_{jk}$.
 Consequently, the components of the tensor $K^i{}_{jk}$ have to be
 linear in $\oG^i{}_{jk}$. Write a general expression of such a type:
  \begin{equation}\label{GCC3}
 K^i{}_{jk}=\frac 12\,\chi_{jkl}{}^{imn}\oG^l{}_{mn}\,.
 \end{equation}
 Since, under a
 transformation of coordinates, the connection $\oG^l{}_{mn}$
 changes by inhomogeneous rule, it can appear in the tensor $K^i{}_{jk}$
 only in the antisymmetric combination. Thus the most general expression for this tensor is
   \begin{equation}\label{GCC4}
K^i{}_{jk}=\frac 12\,\chi_{jkl}{}^{imn}\oG^l{}_{[mn]}=
\frac 12\,\chi_{jkl}{}^{imn}C^l{}_{mn}\,.
\end{equation}
 Hence, the  symmetry relation
$\chi_{jkl}{}^{imn}=\chi_{jkl}{}^{i[mn]}$
 holds.
 The coefficients $\chi_{jkl}{}^{imn}$  have to be constructed from
 the components of the absolute basis $\vt^\a{}_m$ and $e_\a{}^m$.
 Again, since $\chi_{jkl}{}^{imn}$  involves only coordinate indices,
 it has to be constructed from the traced products of the frame and the
 coframe components.
 However all such products are equal to
 the Kronecker symbol. Thus  $\chi_{jkl}{}^{imn}$ has to be  a
 tensor expressed only by the Kronecker symbols.  Consequently, the general expression for
 $\chi_{jkl}{}^{imn}$ can be written as
\begin{equation}\label{GCC5}
\chi_{jkl}{}^{imn}=\a_1\d_j^{[m}\d_k^{n]}\d_l^i+
\a_2\d_l^{[m}\d_k^{n]}\d_j^i+
\a_3\d_l^{[m}\d_j^{n]}\d_k^i\,.
\end{equation}
 Substituting into (\ref{GCC4})  we have
 \begin{equation}\label{GCC6}
 K^i{}_{jk}=\a_1C^i{}_{jk}+\a_2C_j\d^i_k+
 \a_3C_k\d^i_j\,.
\end{equation}
 Consequently  (\ref{GCC1}) is proved.
  $\blacksquare$
%----------------------------
 % end proof

 By (\ref{cc2}), the connection 1-form corresponded to the
 coefficients  (\ref{GCC1}), being referred to a nonholonomic  basis,
 takes the form
 \begin{equation}\label{GCC7}
 \GG_a{}^b={f_a}^k\left(-{\theta^b}_{k,m}+{\theta^b}_l\oG^l{}_{mk}+
 K^l{}_{mk}{\theta^b}_l\right)dx^m\,.
 \end{equation}
When this quantity is referred to the coframe field itself,
the first two terms are canceled.
In this special basis,  the expression is simplified  to
\begin{equation}\label{GCC8}
 \GG_\a{}^\b= K^i{}_{jk}{e_\a}^k{\vt^\b}_idx^j \,,
 \end{equation}
 where $K^i{}_{jk}$ is given in (\ref{GCC6}).
%
% Note that  Weitzenb\"{o}ck's connection appears in this family
% as a trivial case $K^i{}_{jk}$ or, equivalently,  $\a_1=\a_2=\a_3=0$.
 Since the 1-form (\ref{GCC8}) depends only on  antisymmetric
 combinations of the first order derivatives, it can be  expressed by
 the exterior derivative of the coframe:
 \begin{equation}\label{GCC9}
 \GG_\a{}^\b=\left(\a_1C^\b{}_{\g\a}+\a_2C_\g\d^\b_\a+\a_3C_\a\d^\b_\g\right)\vt^\g\,.
 \end{equation}
 Also a components free expression is available
 \begin{equation}\label{GCC9x}
\GG_\a{}^\b=-\frac 12 \left[\a_1e_\a\rfloor d\vt^\b+
 \a_2(e_\a\rfloor \A)\vt^\b + \a_3\d^\b_\a \A\right]\,.
 \end{equation}

%-----------------------------------------

 \vspace{0.3cm}

\noindent{\bf{Metric-coframe connection. }}
 %-----------------------------------------
Consider a manifold endowed with the coframe metric tensor  (\ref{Metrtensor2}).
Again, we are looking for a most general coframe connection that can be constructed from
 the first order derivatives of the coframe field. We will refer to it as the {\it metric-coframe connection}.
 Thus we are deal with a category of  {\it coframe manifolds with a coframe metric and a linear coframe connection:}
 \begin{equation}\label{mcc0}
 \Big{\{}M\,, \vt^\a,G\,,g(\vt^a),\G_a{}^b(\vt^\a)\Big{\}}\,.
 \end{equation}
Now the connection  expression will involve some additional terms which depend on
 the metric tensor (\ref{Metrtensor2}). To describe all possible combinations of the metric tensor components and frame/coframe components it is useful to pull down all the indices.
 Define:
 \begin{equation}\label{mcc1}
 \G_{ijk}=g_{im}\G^m{}_{jk}\,,\qquad C_{ijk}=g_{im}C^m{}_{jk}\,.
 \end{equation}

 \vspace{0.3cm}

 \noindent {\bf Proposition 4:}
 {\it The most general metric-coframe
 connection constructed from the first order derivatives of the
 coframe field is represented by a 6-parametric family:
\begin{equation}\label{mcc5}
\G_{ijk}=\,\oG_{ijk}+\a_1C_{ijk}+\a_2g_{ik}C_j+\a_3g_{ij}C_k+
\b_1g_{jk}C_i+\b_2C_{jki}+\b_3C_{kij}\,.
\end{equation}
}
 \noindent {\bf Proof: }
 Similarly to the case of  a pure coframe connection,  a
 metric-coframe connection  can be represented
 as the Weitzenb\"{o}ck connection plus an
 arbitrary tensor. So  we can write
 \begin{equation}\label{mcc6}
 \G_{ijk}=\,\oG_{ijk}+K_{ijk}\,.
 \end{equation}
 The  tensor $K_{ijk}$ has to be proportional to the
 derivatives of the coframe field $\vt^\a{}_{i,j}$.
Repeating the consideration given above we come to the same conclusion:
the first order derivatives  of the coframe field can appear in the tensor $K_{ijk}$ only via the antisymmetric combination of the flat connection $\oG_{l[mn]}=C_{lmn}$. Consequently we have a relation
\begin{equation}\label{cc14}
 K_{ijk}=\frac 12\,\chi_{ijk}{}^{lmn}C_{lmn}\,.
 \end{equation}
 The tensor $\chi_{ijk}{}^{lmn}$ may involve now the
 components of the metric tensor  in addition to the
 Kronecker symbol. Using the symmetry relation
 $\chi_{ijk}{}^{lmn}=\chi_{ijk}{}^{l[mn]}$
  we construct a most general expression of such a type
 \begin{eqnarray}\label{mcc8}
 \chi_{ijk}{}^{lmn}&=&\a_1\d_i^l\d_j^{[m}\d_k^{n]}+
 \b_2\d_j^l\d_k^{[m}\d_i^{n]}+\b_3\d_k^l\d_i^{[m}\d_j^{n]}+\nonumber\\
 &&\a_2g_{ik}g^{l[m}\d^{n]}_j+\a_3g_{ij}g^{l[m}\d^{n]}_k+
 \b_1g_{jk}g^{l[m}\d^{n]}_i\,.
 \end{eqnarray}
  Consequently, the additional tensor takes the required form
 \begin{eqnarray}\label{mcc9}
 K_{ijk}=\a_1C_{ijk}+\a_2g_{ik}C_j+\a_3g_{ij}C_k+
\b_1g_{jk}C_i+\b_2C_{jki}+\b_3C_{kij}\,.
 \end{eqnarray}
  $\blacksquare$\\
The expression (\ref{mcc5})  can be rewritten in a
\begin{equation}\label{mcc5x}
\G^i{}_{jk}=\,\oG^i{}_{jk}+\a_1C^i{}_{jk}+\a_2\d^i_kC_j+\a_3\d^i_jC_k+
\b_1g^{il}g_{jk}C_l+\b_2g^{il}C_{jkl}+\b_3g^{il}C_{klj}\,.
\end{equation}
In fact, this expression is a proper form of the coefficients of the coframe connection.
Here we can identify two groups of terms: (i) The terms with the coefficient $\a_i$ that do not depend on the metric; (ii) The terms with the coefficient $\b_i$ that can be constructed only by use of the metric tensor.

 With respect to a nonholonomic  basis $(f_a,\theta^a)$,
 the coefficients of a connection (\ref{mcc5x}) correspond to a connection 1-form (\ref{cc2})
 \begin{equation}\label{mcc10}
 \GG_a{}^b=%\left(-{\psi^\ta}_{k,m}+\G_{bc}{}^a{\psi^\ta}_a\right)
 %{f_\tb}^cdx^b
 \oGG_a{}^b+ K^i{}_{jk}{f_a}^k{\theta^b}_idx^j\,.
 \end{equation}
When (\ref{mcc10}) is  referred to the coframe field itself, it is simplified
to
 \begin{equation}\label{mcc11}
\GG_\a{}^\b= K^i{}_{jk}{e_\a}^k{\vt^\b}_idx^j\,.
\end{equation}
This expression depends only
 on the antisymmetric combinations of the first order derivatives
 of the coframe components. So it can be  expressed by the exterior
 derivative of the coframe. We have
 \begin{equation}\label{mcc12}
 \GG_\a{}^\b=\left(\a_1C^\b{}_{\g\a}+\a_2C_\g\d^\b_\a+\a_3C_\a\d^\b_\g+
\b_1C^\b\eta_{\a\g}+\b_2C_{\g\a\nu}\eta^{\b\nu}+\b_3C_{\a\nu\g}\eta^{\b\nu}\right)\vt^\g\,,
  \end{equation}
  or, equivalently,
   \begin{eqnarray}\label{mcc12x}
\GG_\b{}^\a&=&-\frac 12\, \Big[\a_1e_\b\rfloor d\vt^\a+\a_2\vt^\a
 (e_\b\rfloor \A)+\a_3\d^\a_\b \A+\b_1(e^\a\rfloor A)\vt_\b+\nonumber\\
&&\qquad \b_2 e^\a\rfloor (e_\b\rfloor d\vt_\mu)\vt^\mu+
 \b_3e^\a\rfloor d\vt_\b\Big]\,.
 \end{eqnarray}
 %-------------------------------------------------------
 \subsection{Torsion of the coframe connection}
%-------------------------------------------------------
 %-------------------------------------------------------

 \vspace{0.3cm}

\noindent{\bf{Torsion tensor and torsion 2-form. Definitions.}}
%-------------------------------------------------------
 Consider a  connection 1-form $\G_b{}^a$ referred to an arbitrary
 basis $(\theta^a, f_a)$.  For a tensor valued $p$-form
 of a representation type $\rho\left(A_a{}^b\right)$, the {\it covariant exterior
 derivative} operator $D:\Omega^p(\M)\to \Omega^{p+1}(\M)$ is defined as
 \cite{Mielke:1992te}, \cite{Hehl:1994ue}
 \begin{equation}\label{torten1}
 D=d+\GG_b{}^a \rho\left(A_a{}^b\right)\wedge\,.
 \end{equation}
 In particular, the covariant exterior derivative of a scalar-valued
 form $\phi$ is $D\phi=d\phi$. For a vector-valued form $\phi^a$, it
 is given by $D\phi^a= d\phi^a+\GG_b{}^a\wedge\phi^b$, etc.

 For a connection 1-form $\GG_a{}^b$ written with respect to a nonholonomic basis,
 the {\it torsion 2-form} $\T^a$  is  defined as
 \begin{equation}\label{cc22}
 \T^a=D\theta^a=d\theta^a+\GG_b{}^a\wedge\theta^b\,.
 \end{equation}
 On a $D$ dimensional manifold, this  covector valued 2-form has $D(D^2-D)/2$ independent
 components.
 Substituting (\ref{cc2}) into (\ref{cc22}), we observe that the coframe
 derivative term $d\vt^a$ cancels out. Hence,
 \begin{equation}\label{cc23}
 \T^a=\G^i{}_{jk}{}\theta^a{}_idx^j\wedge dx^k=
 \G^i{}_{[jk]}{}\theta^a{}_idx^j\wedge dx^k\,.
 \end{equation}
 In a coordinate coframe, this expression is simplified to
 \begin{equation}\label{cc24}
 \T^i=\G^i{}_{[jk]}{}dx^j\wedge dx^k\,.
 \end{equation}
 Consequently, the torsion 2-form $\T^a$ is completely determined by an antisymmetric combination of the coefficients of the connection. Observe that such combination is a tensor. Thus, the torsion 2-form is completely equivalent to a $(1,2)$-rank {\it torsion tensor} which is defined as
 \begin{equation}\label{cc25}
 T^i{}_{jk}=2\G^i{}_{[jk]}\,.
 \end{equation}
In a holonomic and a nonholonomic bases, the torsion 2-form is expressed respectively as
 \begin{equation}\label{torform}
 \T^i=\frac 12 T^i{}_{jk}{}dx^j\wedge dx^k\,,\qquad \T^a=\frac 12 T^i{}_{jk}{}\theta^a{}_idx^j\wedge dx^k \,.
 \end{equation}
It is useful to define also a quantity
\begin{equation}\label{torform1}
 \T^\a=\frac 12 T^i{}_{jk}{}\vt^\a{}_idx^j\wedge dx^k\,.
 \end{equation}
 Observe that this set of  2-forms cannot be regarded as a vector-valued form since the transformations of the coframe field $\vt^\a$ are restricted.
 However, the proper vector valued torsion 2-forms (\ref{torform}) are related to the quantity (\ref{torform1}) by the following simple equations
 \begin{equation}\label{torform2}
\T^i=e_\a{}^i\T^\a\,,\qquad \T^a=\theta^a{}_ie_\a{}^i\T^\a \,.
 \end{equation}
With respect to the coframe field, the torsion 2-form of the Weitzenb\"{o}ck connection (\ref{torform1}) reads
\begin{equation}\label{cc31x}
 \oTT^\a=d\vt^\a\,.
\end{equation}

%-------------------------------------------------------

 \vspace{0.3cm}

\noindent{\bf{Torsion of the metric-coframe connection. }}
%-------------------------------------------------------
 For the metric-coframe connection (\ref{mcc5}), the covariant components $T_{ijk}=2g_{im}\G^m{}_{[jk]}$  of the torsion tensor take the form
 \begin{equation}\label{mcc13}
 T_{ijk}=2(1+\a_1)C_{ijk}+(\a_2-\a_3) (g_{ik}C_j-g_{ij}C_k)+(\b_2+\b_3)
(C_{jki}+C_{kij})\,.
 \end{equation}
 The corresponded torsion 2-form is expressed in the coordinate basis as
 \begin{equation}\label{mcc13x}
 T^i=\Big[(1+\a_1)C^i{}_{jk}+(\a_2-\a_3) C_j\d^i_k+(\b_2+\b_3)
g^{im}C_{jkm}\Big]dx^j\wedge dx^k\,.
 \end{equation}
 %Using the definition $T^\a=d\vt^\a+\G_\b{}^\a\wedge \vt^\b$, we derive
% \begin{equation}\label{mcc20}
 %\T^\a=(1+\a_1)d\vt^a-\frac 12\, (\a_2-\a_3)\vt^\a\wedge \A+\frac 12\,
 %(\b_2-\b_3)\vt^\mu\wedge(e^\a\rfloor d\vt_\mu)\,,
 % \end{equation}
 %or, equivalently,
 In term of the differential forms $\A$ and $\B$ (see Appendix) we derive
 \begin{equation}\label{mcc21}
 \T^\a=\left(1+\a_1\right)d\vt^a-\frac 12\,
 (\a_2-\a_3)\vt^\a\wedge \A-\frac 12\, (\b_2+\b_3)\Big(d\vt^a-e^\a\rfloor \B\Big)\,.
  \end{equation}

 %-----------------------

 \vspace{0.3cm}

\noindent{\bf{Irreducible decomposition of the torsion. }}
 %-----------------------
On a manifold of a dimension $D\ge3$ endowed with a metric
 tensor, the
 torsion 2-form admits an irreducible decomposition into three
 independent pieces \cite{Hehl:1994ue}
 \begin{equation}\label{mcc22}
 \T^a={}^{(1)}\T^a+{}^{(2)}\T^a+{}^{(3)}\T^a\,.
 \end{equation}
 Here {\it the trator} and {\it the axitor} parts \cite{Hehl:1994ue} are defined correspondingly as
  \begin{equation}\label{mcc23}
 {}^{(2)}\T^a=\frac 1{n-1}\,\theta^a\wedge (f_b\rfloor \T^b)\,,\qquad
 {}^{(3)}\T^a=\frac 13 f^a\rfloor (\theta^b\wedge \T_\b)\,.
 \end{equation}
 The remainder ${}^{(1)}\T^a$ is referred to as a {\it tentor part}.
 The irreducible decomposition means that the different pieces transform
 independently by the same tensorial rule as the total quantity.
 Particularly, we can check straightforwardly that for every part of the torsion tensor
  \begin{equation}\label{mcc26}
 {}^{(p)}\T^a={}^{(p)}\T^\a\theta^a{}_i e_\a{}^i \,, \qquad p=1,2,3.
 \end{equation}
 So it is enough to provide the calculations of the irreducible pieces with respect to the coframe field itself.
  We have the second piece of the torsion as
 \begin{equation}\label{mcc27}
 {}^{(2)}\T^\a=\frac 1{n-1}\,\vt^\a\wedge (e_\b\rfloor \T^\b)
 = \frac {\tau_2}{2(n-1)}\, \vt^\a\wedge \A\,,
 \end{equation}
 where
 \begin{equation}\label{mcc28}
 \tau_2=2(1+\a_1)-(\b_2+\b_3)-(\a_2-\a_3)(n-1)\,.
 \end{equation}
 The third piece of torsion is given by
 \begin{equation}\label{mcc29}
 {}^{(3)}\T^\a=\frac 13 e^\a\rfloor(\vt^\b\wedge \T_\b)
 = \frac {\tau_3}3\, e^\a\rfloor \B\,,
 \end{equation}
 where
 \begin{equation}\label{mcc30}
 \tau_3=(1+\a_1)+(\b_2+\b_3)\,.
 \end{equation}
 The first part takes the form
 \begin{eqnarray}\label{mcc31}
 {}^{(1)}\T^\a&=&\T^\a-{}^{(2)}\T^\a-{}^{(3)}\T^\a %\nonumber\\
 %&=&
=\tau_1 \left( d\vt^a-\frac
 1{n-1}\, \vt^\a\wedge \A-\frac 13\, e^\a\rfloor \B\right)\,,
 \end{eqnarray}
 where
 \begin{equation}\label{mcc32}
 \tau_1=(1+\a_1)-\frac 12 (\b_2+\b_3)\,.
 \end{equation}

%-------------------------------------------------------
 
 \vspace{0.3cm}

\noindent{\bf{Torsion-free metric-coframe connection. }}
%-------------------------------------------------------
 Let us look for which values of the parameters the torsion
 of the metric-coframe connection is identically zero.
 The corresponded connection is called {\it the symmetric or torsion-free
 connection}.
It is clear from (\ref{mcc13}) that   the metric-coframe connection is
 symmetric if
 \begin{equation}\label{mcc14}
 \a_1=-1\,,\qquad \a_2=\a_3\,,\qquad \b_2=-\b_3\,.
 \end{equation}
 The necessity of this condition can be derived from the irreducible decomposition.
 Indeed, since the three pieces of the torsion are mutually independent, they have to vanish simultaneously. Hence we have a condition $\tau_1=\tau_2=\tau_3=0$ which is equivalent to
 (\ref{mcc14}).
 Note that this requirement is necessary only for a manifold of the dimension $D\ge 3$. On a two-dimensional manifold, the metric-coframe connection is symmetric under a weaker condition
 \begin{equation}\label{mcc16}
 2(1+\a_1)+(\a_2-\a_3)- (\b_2+\b_3)=0\,.
 \end{equation}
 On a curve, every  connection is unique and
 symmetric.

Thus on a manifold of the dimension $D\ge 3$ there exists a 3-parametric family of the symmetric (torsion-free) connections:
 \begin{eqnarray}\label{mcc15x}
\G^i{}_{jk}=\,\oG^i{}_{jk}-C^i{}_{jk}+\a_2\left(\d^i_kC_j+\d^i_jC_k\right)+
\b_1g_{jk}g^{im}C_m+\b_2g^{im}\left(C_{jkm}-C_{kmj}\right)\,.
 \end{eqnarray}
%--------------------------------
\subsection{Nonmetricity of the metric-coframe connection}
 %--------------------------------
 
 \vspace{0.3cm}

\noindent{\bf{Nonmetricity tensor and nonmetricity 2-form. Definition. }}
 When  Cartan's manifold is  endowed with a metric tensor, the  connection
 generates an additional tensor field called {\it the nonmetricity
 tensor}. It is expressed as a covariant derivative of the metric
 tensor components. For  a metric  given in a local system of
 coordinates as $g=g_{ij}dx^i\otimes dx^j$, the nonmetricity
 tensor is defined as 
 \begin{equation}\label{nonmetr1}
 Q_{kij}=-\nabla_kg_{ij}=
 -g_{ij,k}+\G^m{}_{ik}g_{mj}+\G^m{}_{jk}g_{im}\,,
  \end{equation}
  or,
  \begin{equation}\label{nonmetr1x}
 Q_{kij}=-g_{ij,k}+\G_{jik}+\G_{ijk}\,.
 \end{equation}
 Evidently, this tensor is symmetric in the last  pair of indices
 $Q_{kij}=Q_{kji}$. Hence, on a $D$ dimensional manifold, the nonmetricity tensor has  $D(D^2+D)/2$
 independent components.

 For the exterior form representation, it is useful to define {\it the nonmetricity 1-form}.
 In a coordinate basis, it is given by
 \begin{equation}\label{nonmetr2}
 Q_{ij}=Q_{kij}dx^k= -dg_{ij}+\G_{ij}+\G_{ji}\,.
 \end{equation}
 In an arbitrary reference basis $\left(f_a\,,\theta^a\right)$, the metric
 tensor is expressed as $g=g_{ab}\theta^a\otimes\theta^b$.
 Correspondingly, the nonmetricity 1-form  reads
  \begin{equation}\label{mcc35}
 Q_{ab}=-dg_{ab}+\G_{ab}+ \G_{ba}\,.
 \end{equation}
 With respect to the coframe field $\vt^\a$, the components of the metric are
 constants $\eta_{\a\b}$, thus the nonmetricity is merely the symmetric
 combination of the connection 1-form components
 \begin{equation}\label{mcc36}
 Q_{\a\b}=\G_{\a\b}+\G_{\b\a}\,.
 \end{equation}
 Note, that this expression is not a usual tensorial quantity. In fact, it is an expression of a tensor-valued 1-form of nonmetricity with respect to a special class of bases.
 Its relation to a proper tensorial valued 1-form  (\ref{nonmetr2}) is, however, very simple.
By a substitution of (\ref{cc5}) into (\ref{mcc36}) we have
 \begin{equation}\label{mcc36x}
 Q_{ij}=Q_{\a\b}\vt^{\a}{}_i\vt^{\b}{}_j\,.
 \end{equation}

%\sub\\sub\section{Decomposition of an affine connection}
 The following generalization of the Levi-Civita theorem from the  Riemannian geometry provides a decomposition of an arbitrary affine connection \cite{Schou}. Its simple proof  is instructive for our construction.

 \vspace{0.3cm} \noindent {\bf Proposition 5:}
\noindent  {\it  Let a metric $g$ on a manifold M be fixed and two
 tensors $T_{ijk}$ and $Q_{ijk}$ with the symmetries
 \begin{equation}\label{mcc37}
 T_{ijk}=-T_{ikj}\,, \qquad Q_{kij}=Q_{kji}\,.
 \end{equation}
 be given. A unique connection $\G_{ijk}$ exists  on $M$
 such that $T_{ijk}$ is its torsion and $Q_{ijk}$ is its
 nonmetricity.
 Explicitly,
 \begin{eqnarray}\label{mcc38}
 \G_{ijk}&=&\LG_{ijk}-
 \frac 12 \big(Q_{ijk}-Q_{jki}-Q_{kij}\big)+\frac 12\big(T_{ijk}+T_{jki}-
 T_{kij}\big)\,,
 \end{eqnarray}
 where
  \begin{equation}\label{mcc38x}
\LG_{ijk}=\frac12\big(g_{ij,k}+g_{ik,j}-g_{jk,i}\big)
 \end{equation}
 are the components of the Levi-Civita connection.
 }

\vspace{0.3cm}

\noindent {\bf Proof: }
On a $D$-dimensional manifold  definitions of the torsion and the nonmetricity tensors
 \begin{equation}\label{proof1}
T_{ijk}=2\G_{i[jk]}\,, \qquad Q_{kij}=-g_{ij,k}+\G_{ijk}+\G_{jik}
 \end{equation}
can be viewed as a linear system of $D^3$  linear equations for $D^3$ independent
 variables $\G_{ijk}$
 \begin{equation}\label{proof2}
 \G_{i[jk]}=\frac 12T_{ijk}\,,\qquad
 \G_{(ij)k}=\frac 12 (Q_{kij}+g_{ij,k})\,.
 \end{equation}
%This is a system of $n^3$ linear equations for $n^3$ independent variables $\G_{ijk}$
%\[\frac{n(n^2+n)}2+\frac{n(n^2-n)}2=n^3\]
 For
$T_{ijk}=Q_{kij}=0$, the system has a unique solution --- the
Levi-Civita connection $\LG_{ijk}$. Thus the determinant of the
matrix of the system (\ref{proof2}) is nonsingular.
Consequently also for arbitrary tensors $T_{ijk}$ and $Q_{kij}$,
the system has a unique solution. In order to check the specific form of
the solution (\ref{mcc38}), it is enough to substitute  the
definitions (\ref{proof1}). $\blacksquare$

%-------------------------------------

 \vspace{0.3cm}

\noindent{\bf{Nonmetricity of the metric-coframe connection. }}
%-------------------------------------
 We calculate now the nonmetricity tensor of the metric-coframe connection (\ref{mcc5})
 \begin{eqnarray}\label{mcc39}
 Q_{kij}&=&\Big(-g_{ij,k}+\oG_{ijk}+\oG_{jik}\Big)+\nonumber\\
 &&(\a_1-\b_2)(C_{ijk}-C_{jki})+(\a_2+\b_1)(g_{ik}C_j+g_{jk}C_i)+2\a_3g_{ij}C_k\,.
 \end{eqnarray}
 The first parenthesis represent the nonmetricity tensor of the
  Weitzenb\"{o}ck connection. This expression vanishes identically,
  i.e., the Weitzenb\"{o}ck connection is metric-compatible.
 Indeed, we have
 \begin{equation}\label{mcc40}
 g_{ij,k}=\eta_{\a\b}(\vt^\a{}_{i,k}\vt^\b_j+\vt^\a{}_{i}\vt^\b_{j,k})
 =\oG_{ijk}+\oG_{jik}\,.
 \end{equation}
 Consequently, (\ref{mcc39})  is simplified to
 \begin{equation}\label{mcc41}
 Q_{kij}=(\a_1-\b_2)(C_{ijk}+C_{jki})+
 (\a_2+\b_1)(g_{ik}C_j+g_{jk}C_i)+2\a_3g_{ij}C_k\,.
 \end{equation}
Relative to the coframe field, we have, using (\ref{mcc12},\ref{mcc36}),
 the 1-form of nonmetricity
 \begin{eqnarray}\label{mcc42}
 Q_{\a\b}&=&-\frac 14\, (\a_1-\b_2)\Big(e_\a\rfloor d\vt_\b+
 e_\b\rfloor d\vt_\a\Big)+
 \frac 12 \, \a_3\eta_{\a\b}A+\nonumber\\
 && \frac 14\, (\a_2+\b_1)\Big[(e_\a\rfloor A)\vt_\b+
 (e_\b\rfloor A)\vt_\a\Big]\,.
 \end{eqnarray}
 %-----------------------------------------
 
 \vspace{0.3cm}

\noindent{\bf{Irreducible decomposition of the nonmetricity.}}
 %-----------------------------------------
 We are looking now for an irreducible decomposition of the
 nonmetricity 1-form $Q_{ab}$ under the
 pseudo-orthogonal group.
 Since $Q_{ab}$ is a tensor-valued 1-form it can be
 calculated in an arbitrary basis. Certainly, the  basis of the
 coframe field is
 the best for these purposes. We have only remember that for a
 transformation to an arbitrary basis we have simple multiply
 the corresponding quantity $Q_{\a\b}$ by the matrix of the transformation.
 We cannot, however, transform the coframe basis to an arbitrary
 basis. This is because the coframe field is a fixed building block of
 our construction.

 The irreducible decomposition of the nonmetricity 1-form under the
 pseudo-orthogonal group $SO(1,n)$ is constructed by the in
 correspondence to the Young diagrams.
 For actual calculations we use the  algorithm given in
 \cite{Hehl:1994ue}.
 The resulting decomposition is given as a sum of four independent pieces
 \begin{equation}\label{mcc49}
 Q_{\a\b}={}^{(1)}Q_{\a\b}+{}^{(2)}Q_{\a\b}+
 {}^{(3)}Q_{\a\b}+{}^{(4)}Q_{\a\b}\,.
 \end{equation}
For the   nonmetricity  1-form (\ref{mcc36}),the irreducible parts are
\begin{equation}\label{mcc50}
 {}^{(1)}Q_{\a\b}=\mu_1 \Big[(n-1)e_{(\a}\rfloor d\vt_{\b)}+
 (e_{(\a}\rfloor A)\vt_{\b)}-4\eta_{\a\b}A\Big]\,,
 \end{equation}
\begin{equation}\label{mcc51}
 {}^{(2)}Q_{\a\b}=\mu_2 \Big[(n-1)e_{(\a}\rfloor d\vt_{\b)}+
 (e_{(\a}\rfloor A)\vt_{\b)}+ 2\eta_{\a\b}A\Big]\,,
 \end{equation}
\begin{equation}\label{mcc52}
 {}^{(3)}Q_{\a\b}=\mu_3 \Big[(e_{(\a}\rfloor A)\vt_{\b)}+
 \frac 2n\,\eta_{\a\b}A\Big]\,,
 \end{equation}
\begin{equation}\label{mcc53}
 {}^{(4)}Q_{\a\b}=\mu_4 \Big[\frac 1n\,\eta_{\a\b}A\Big]\,.
 \end{equation}
 The coefficients of these quantities depend on the parameters  of
 the general connection as
\begin{eqnarray}\label{mcc54}
\mu_1&=&- \, \frac 1{6(n-1)}\,(\a_1-\b_2) \,,\qquad \mu_2=\frac
1{2} \, \mu_1\,,\\ \label{mcc55}
 \mu_3&=&\frac 14 \,
\Big[\frac 1{n-1}\, (\a_1-\b_2)+(\a_2+\b_1)\Big]\,,\\
\label{mcc56} \mu_4&=&\frac 12\,
\Big[-(\a_1-\b_2)+n\a_3+(\a_2+\b_1)\Big]\,.
 \end{eqnarray}
% Again, the conditions $\mu_1=\mu_2=\mu_3=\mu_4=0$ are
% eqivalent to (\ref{mcc43}).
 
 \vspace{0.3cm}

\noindent{\bf{Metric compatible metric-coframe connection.}}
 Let us look for which values of the coefficients the connection is
 {\it metric-compatible}, i.e., has an identically zero non-metricity
 tensor. Recall that both quantities, the metric tensor and the
 connection, are constructed from the same building block --- the
 coframe field $\vt^\a$.
It is clear from (\ref{mcc42}) that   the metric-coframe
connection is metric-compatible if
 \begin{equation}\label{mcc43}
 \a_1=\b_2\,,\qquad \a_2=-\b_1\,,\qquad \a_3=0\,.
 \end{equation}
 The necessity of this condition can be derived from the irreducible
 decomposition of the nonmetricity tensor.
 Four irreducible  pieces of the non-metricity tensor are mutually
 independent, so they have to vanish simultaneously.
 Hence we have a condition $\mu_1=\mu_2=\mu_3=\mu_4=0$
 which turns out to be  equivalent to (\ref{mcc43}).
 Note that this requirement is necessary only for a manifold
 of the dimension $D\ge 3$, where the irreducible decomposition
 (\ref{mcc49}) is valid.
 On a two-dimensional manifold, the metric-coframe connection is
 metric-compatible if and only if
 \begin{equation}\label{mcc44}
 \a_1-\b_2=\a_2+\b_1=\a_3\,.
 \end{equation}
 On a one-dimensional manifold, every  connection is metric-compatible.

 \vspace{0.3cm}

\noindent{\bf{Metric compatible and torsion-free metric-coframe connection.}}
 Let us  look now for a general coframe connection of a zero
 torsion and zero non-metricity, i.e., for a symmetric metric
 compatible connection constructed from the coframe field.
 The system of conditions (\ref{mcc14}) and (\ref{mcc43}) has a
 unique solution
   \begin{equation}\label{mcc46}
 \a_1=\b_2=-\b_3=-1\,,\qquad \b_1=\a_2=\a_3=0\,.
 \end{equation}
 Consequently, a metric-compatible symmetric connection is unique.
 This is in a correspondence to the original Levi-Civita theorem, and the
 unique connection is of Levi-Civita.
 Moreover, substituting (\ref{mcc46}) into (\ref{mcc5}) we can express
 now the standard Levi-Civita connection $ \LG^i{}_{jk}$ via
 the flat connection of Weitzenb\"{o}ck  $ \oG^i{}_{jk}$ ---
  \begin{eqnarray}\label{mcc47}
 \LG{ijk}&=&\oG_{i(jk)}+C_{kij}-C_{jki}\,.
 \end{eqnarray}
 In the  basis constructed from the coframe field itself,
 the nonmetricity 1-form   for the Levi-Civita
 connection reads
 \begin{eqnarray}\label{mcc48}
\LG_{\a\b}&=&e_\a\rfloor d\vt^\b- e_\b\rfloor d\vt^\a-
 \frac 12\,  e_\a\rfloor e_\b\rfloor B\,.
\end{eqnarray}
It is in a correspondence with a formula given in
\cite{Hehl:1994ue}.
 %@@@ UP TO HERE!!!
%Calculating the following expression
%\begin{eqnarray}\label{mcc49}
% e_\b\rfloor (e^\a\rfloor B)&=&e_\b\rfloor\Big[(e^\a\rfloor d\vt_\mu)\wedge \vt^\mu+d\vt^\mu\Big]
% \nonumber\\
% &=&\Big[e_\b\rfloor(e^\a\rfloor d\vt_\mu)\Big]\wedge \vt^\mu-e^\a\rfloor d\vt_\b+e_\b\rfloor d\vt^\a
% \end{eqnarray}
% we have a compact expression for the Levi-Civita connection.
%
%\vspace{0.3cm} \noindent {\bf Theorem 9}
%
%\noindent {\it In the absolute orthogonal basis, the  Levi-Civita connection 1-form is given as
%  \begin{equation}\label{ngc18}
% \LG_\b{}^\a=\frac 12 \, e_\b\rfloor e^\a\rfloor B\,,
% \end{equation}
% where $B=d\vt^\mu\wedge \vt_\mu$.
% }
%
%\vspace{0.3cm}

 %----------------------------------------------
 \subsection{Gauge transformations}
%----------------------------------------------
 
 \vspace{0.3cm}

\noindent{\bf{Local transformations of the coframe field.}}
 %----------------------------------------------
The geometrical structure considered above is well defined for a fixed
 coframe field $e_\a$.
 Moreover, it is invariant under  rigid coframe transformations.
 The gauge paradigm suggests now to look for
 a localization of such transformations:
\begin{equation}\label{gt1}
 \vt^\a\mapsto L^\a{}_\b\,\vt^\b\,,\qquad e_\a\mapsto
 L_\a{}^\b \,e_\b\,,
\end{equation}
 or, in the components,
\begin{equation}\label{gt1-1}
 \vt^\a{}_i\mapsto L^\a{}_\b\,\vt^\b{}_i\,,\qquad e_\a{}^i\mapsto
 L_\a{}^\b{} \,e_\b{}^i\,.
 \end{equation}
 Here the matrix $L^\a{}_\b$ and its inverse $L_\a{}^\b$ are functions of a point $x\in M$.
We require the volume element
(\ref{volel1}) and the metric tensor (\ref{Metrtensor2}) both to
be invariant under the pointwise transformations (\ref{gt1}).
Consequently, $L^\a{}_\b$ is assumed to be a pseudo-orthonormal
matrix whit enters are smooth functions of a point. We will also use an
 infinitesimal version of the
 transformation (\ref{gt1-1}) with $L^\a{}_\b=\d^\a_\b+X^\a{}_\b$.
 In the components, it takes the form
 \begin{equation}\label{gt2}
 \vt^\a{}_i\mapsto \vt^\a{}_i+X^\a{}_\b\,\vt^\b{}_i\,,\qquad
 e_\a{}^i\mapsto e_\a{}^i-X^\b{}_\a\, e_\b{}^i\,.
\end{equation}
As the elements of the algebra $so(1,n)$, the matrix $X_{\a\b}=\eta_{\a\mu}X^\mu{}_\b$ is antisymmetric. We define a corresponded antisymmetric tensor 
\begin{equation}\label{ant-tensors}
F_{ij}=\vt^\a{}_i\vt^\b{}_jX_{\a\b}\,.
\end{equation}

 \vspace{0.3cm}

\noindent{\bf{Connection invariance postulate. }}
 Recall that we are looking for a most general geometric structure
that can be explicitly constructed from the coframe field.
Moreover, we are interested not in a one fixed coframe field, but
rather in a family of fields related by the left action of the
elements of some continuous group $G$.

In a general setting, the different geometrical
structures such as  the volume element, the metric tensor, and the field
of affine connections, are completely independent. We have
already postulated the invariance of the volume element and of the
metric tensor under the coframe transformations. It is natural to involve now an additional
invariance requirement concerning the affine connection.
\begin{itemize}
\item[]{\underline{\it Connection invariance postulate:}}
 Affine coframe connection  is assumed to be invariant under
 pointwise transformations of the coframe field
\begin{equation}\label{cof-invar}
\G^i{}_{jk}\left(\vt^\a\right)=
\G^i{}_{jk}\left(L^{\a}{}_\b\vt^\b\right)\,.
\end{equation}
\end{itemize}
Since the coframe connection is constructed from the first order
derivatives of the coframe field, (\ref{cof-invar}) is a first
order PDE for the elements of the group $G$ and for the
components of the coframe field.
%---------------------------------

 \vspace{0.3cm}

\noindent{\bf{Weitzenb\"{o}ck connection transformation.}}
%---------------------------------
 Since the Weitzenb\"{o}ck connection is a basis
tool of our construction,
 it is useful to calculate the change of this quantity  under the coframe transformations
(\ref{gt1}). We have
 \begin{equation}\label{gt1-2}
 \Delta\oG^i{}_{jk}=e_\a{}^i\vt^\b{}_k Y^\a{}_{\b j}\,, \qquad {\rm where}
 \qquad  Y^\a{}_{\b j}=L^\a{}_\g L^\g{}_{\b,j}\,.
 \end{equation}
 All matrices involved here are nonsingular, consequently the
  Weitzenb\"{o}ck connection is preserved only under the rigid
  transformations of the coframe field with  $L^\g{}_{\b,j}=0$.

  Let us rewrite (\ref{gt1-2}) in alternative forms.
  Since the metric tensor is invariant under the transformations
 (\ref{gt1}) we have
 \begin{equation}\label{gt1-2x}
 \Delta\oG_{ijk}=\Delta\left(g_{im}\oG^m{}_{jk}\right)=
 g_{im}\Delta\oG^m{}_{jk}\,.
 \end{equation}
 Consequently 
  \begin{equation}\label{gt1-2y}
 \Delta\oG_{ijk}=\vt^\a{}_i\vt^\b{}_kY_{\a\b j}\,,\qquad {\rm where}\qquad
 Y_{\a\b j}=\eta_{\a\mu}Y^\mu{}_{\b j}\,.
 \end{equation}
In the infinitesimal approximation, (\ref{gt1-2}) takes the form
 \begin{equation}\label{gt2-2}
 \Delta\oG^i{}_{jk}=e_\a{}^i\vt^\b{}_k X^\a{}_{\b, j}\,.
 \end{equation}
 while (\ref{gt1-2x}) with $X_{\a\b}=\eta_{\a\mu}X^\mu{}_\b$ reads
 \begin{equation}\label{gt2-2x}
 \Delta\oG_{ijk}=\vt^\a{}_i\vt^\b{}_kX_{\a\b,j} \,.
  \end{equation}
  Note that since $X_{\a\b}$ is antisymmetric, we have in this
  approximation
 \begin{equation}\label{gt2-2xx}
 \Delta\oG_{ijk} =-\Delta\oG_{kji}\,.
  \end{equation}
We will also consider an additional physical meaningful approximation
when the derivatives of the coframe is considered to be
small  relative to the
 derivatives of the transformation  matrix.
 In this case, (\ref{gt2-2}) and  (\ref{gt2-2x})  read
 \begin{equation}\label{gt2-3}
 \Delta\oG^i{}_{jk}=F^i{}_{k,j} \,, \qquad {\rm where}
 \qquad  F^i{}_k=e_\a{}^i\vt^\b{}_k X^\a{}_{\b}\,,
 \end{equation}
 and
 \begin{equation}\label{gt2-3x}
 \Delta\oG_{ijk}=F_{ik,j} \,, \qquad {\rm where}
 \qquad  F_{ij}=\vt^\a{}_i\vt^\b{}_j X_{\a\b}\,.
 \end{equation}
 %----------------------------------------------
 
 \vspace{0.3cm}

\noindent{\bf{Transformations preserved the geometric structure.}}
%----------------------------------------------
% We consider now the {\it gauge coframe structure}
% \begin{equation}\label{gt3}
% \{M\,,g(\vt^\a)\,,\G_{ab}{}^c(\vt^\a)\,,G\}\,.
%\end{equation}
%Here $M$ is a differentiable manifold, g is a metric on $M$
 %wich dependence on the coframe is given in (\ref{mcc1}),
 %$\G_{ab}{}^c(\vt^\a)$ is a linear coframe connection (\ref{mcc5}),
 %and $G$ is a group of the transformations of absolute basis (\ref{gt1})
 %which elements are functions of a point $x\in M$.
Since the coframe field appears in the coframe geometrical structure 
 only implicitly, (\ref{gt1})
 is a type of a gauge transformation.
 Invariance of the metric tensor and of the volume element restricts $L^\a{}_\b$ to
 a pseudo-orthonormal matrix $G=SO(1,n)$.
 % In the infinitesimal version,
 %it means that the matrix $X_{\a\b}=X^\mu{}_\b \eta_{\mu\a}$
 %is antisymmetric.
Let us ask now, under what conditions the  general coframe  connection (\ref{mcc5})
  is invariant under the coframe transformations (\ref{gt1}).
 First we rewrite (\ref{mcc5}) via the Levi-Civita connection. Using (\ref{mcc46}) we have 
 \begin{equation}\label{gt4}
 \oG_{ijk}=\LG_{ijk}+C_{ijk}-C_{kij}+C_{jki}\,.
 \end{equation}
 Thus (\ref{mcc5}) takes the form 
 \begin{equation}\label{gt5}
 \G_{ijk}=\,\LG_{ijk}+(\a_1+1)C_{ijk}+\a_2g_{ik}C_j+\a_3g_{ij}C_k+
\b_1g_{jk}C_i+(\b_2+1)C_{jki}+(\b_3-1)C_{kij}\,.
  \end{equation}
 Since the Levi-Civita connection $\LG_{ijk}$ is invariant
 under the transformations (\ref{gt1}),  the equation $\Delta  \G_{ijk}=0$ 
 takes the form
 \begin{equation}\label{gt8}
(\a_1+1)\Delta C_{ijk}+\a_2g_{ik}\Delta C_j+\a_3g_{ij}\Delta C_k+
\b_1g_{jk}\Delta C_i+(\b_2+1)\Delta C_{jki}+(\b_3-1)\Delta C_{kij}=0\,.
  \end{equation}
Hence in order to have an invariant coframe connection, we have to
look for possible solutions of  equation (\ref{gt8}).
%------------------------------------

 \vspace{0.3cm}

\noindent{\bf{Trivial solutions of the invariance equation.}}
%-------------------------------------
Consider first two trivial solutions of (\ref{gt8}) which turn out to be  non-dynamical.

(i) {\it Arbitrary transformations --- Levi-Civita connection.}\\
The equation (\ref{gt8}) is evidently satisfied when all the
numerical  coefficients mutually equal to zero. It is easy to
check that these six relations are equivalent to (\ref{mcc46}).
Thus the corresponded connection is of Levi-Civita.
 In this case, the elements of the matrix $L^\a{}_\b$ are
 arbitrary functions of a point. Thus we come to a trivial fact
 that the Levi-Civita connection is a unique coframe connection
 which is invariant under arbitrary local $SO(1,n)$ transformations of
 the coframe field.

 (ii) {\it Rigid transformations.}\\
 Another trivial solution of the system (\ref{gt8}) emerges when
 we require $\Delta C_{ijk}=0$. All permutations and traces of
 this tensor are also equal to zero so (\ref{gt8}) is trivially
 valid.  Due to
 (\ref{gt1-2y}), it means that the matrix of transformations is
 independent on a point. In this case, an arbitrary coframe connection, in particular
 the Weitzenb\"{o}ck connection, remains unchanged.
 Thus we come to another trivial fact that the coframe connection is
 invariant under rigid transformations of the coframe field.

  %-------------------------------------
 
 \vspace{0.3cm}

\noindent{\bf{Dynamical solution.}}
 We will look now for nontrivial solutions of the system
(\ref{gt8}). Three traces of this system yield the equations of the
type $\lambda \Delta C_i=0$, where $\lambda$ is a linear combination of
the coefficients $\a_i,\b_i$.  Thus we have to apply the first condition
 \begin{equation}\label{gt9}
 \Delta C_i=0\,.
 \end{equation}
The system (\ref{gt8}) remains now in the form
 \begin{equation}\label{gt10}
 (\a_1+1)\Delta C_{ijk}+(\b_2+1)\Delta C_{jki}+(\b_3-1)\Delta C_{kij}=0\,.
 \end{equation}
Applying the complete antisymetrization in three indices 
we derive the second equation
 \begin{equation}\label{gt11}
 \Delta C_{[ijk]}=0\,.
 \end{equation}
The equation (\ref{gt10}) remains now in the form
\begin{equation}\label{gt12}
 (\b_2-\a_1)\Delta C_{jki}+(\b_3-\a_1-2)\Delta C_{jki}=0\,.
 \end{equation}
We have to restrict now the coefficients,  otherwise we obtain 
$\Delta C_{ijk}=0$, i.e., only the rigid transformations.
Consequently we require
\begin{equation}\label{gt12x}
 \b_2=\a_1\,,\qquad \b_3=\a_1+2\,.
 \end{equation}
Thus we have proved

 \vspace{0.3cm} \noindent {\bf Proposition 6:}
\noindent  {\it  The coframe connection 
\begin{equation}\label{gt13}
\G_{ijk}=\,\LG_{ijk}+(\a_1+1)C_{[ijk]}+\a_2g_{ik}C_j+\a_3g_{ij}C_k+
\b_1g_{jk}C_i\,.
 \end{equation}
 is invariant under the coframe transformations satisfied
the equations
\begin{equation}\label{gt14}
\Delta C_i=0\,.\qquad \Delta C_{[ijk]}=0\,.
 \end{equation}
}
%\vspace{0.3cm}
Observe that this family includes the Levi-Civita connection, which is invariant under arbitrary transformations of the coframe field. 
The torsion tensor of the connection (\ref{gt13}) is expressed as 
 \begin{equation}\label{gt14x}
T_{ijk}=(\a_1+1)C_{[ijk]}+(\a_2-\a_3)(g_{ik}C_j-g_{ij}C_k)\,.
 \end{equation}
Thus a torsion-free subfamily of (\ref{gt13}) is given by
\begin{equation}\label{gt15}
\G_{ijk}= \LG_{i(jk)}+
\a_2(g_{ik}C_j+g_{ij}C_k)+\b_1g_{jk}C_i\,.
 \end{equation}
The nonmetricity  tensor of the connection (\ref{gt13}) reads
\begin{equation}\label{gt16}
\Q_{kij}= (\a_2+\b_1)(g_{ik}C_j+g_{jk}C_i)+2\a_3g_{jj}C_k\,.
 \end{equation}
Thus a metric compatible subfamily of (\ref{gt13}) is given by
\begin{equation}\label{gt17}
\G_{ijk}= \LG_{i(jk)}+(\a_1+1)C_{[ijk]}+\a_2(g_{ik}C_j-g_{jk}C_i)
\,.
 \end{equation}
 From (\ref{gt14x}) and (\ref{gt16}) we derive an interesting conclusions: 
 \begin{equation}\label{gt18}
\Delta Q_{kij}=0\quad \Longleftrightarrow \quad \Delta C_i=0\,.
 \end{equation}
 and, together with this relation, 
  \begin{equation}\label{gt19}
\Delta T_{ijk}=0\quad \Longleftrightarrow \quad \Delta C_{[ijk]}=0\,.
 \end{equation}
 Thus the relations (\ref{gt14}) obtain a geometric meaning, they correspond to invariance of the torsion and nonmetricity tensors under coframe transformations.
 %------------------------------
\subsection{Maxwell-type system}
 %------------------------------
 Let us examine  now what physical meaning can be given to the
invariance conditions \cite{Itin:2006pd}
\begin{equation}\label{PreMax-x0}
\Delta C_{[ijk]}=0\,,\qquad  \Delta C_i=0\,.
 \end{equation}
 Denote $K_{ijk}=\Delta C_{ijk}$. Thus (\ref{PreMax-x0}) takes the form
\begin{eqnarray}\label{PreMax-x}
K_{[ijk]}=0\,, \qquad K^m{}_{im}=0\,.
\end{eqnarray}%\labe{PreMax-x}
The tensor $ K_{ijk}$ depends on the derivatives of the Lorentz
parameters $X_{\a\b}$ and on the components of the coframe field
\begin{equation}\label{KK-def}
 K_{ijk}=\frac 12 \, \vt^\a{}_k\Big(X_{\a\b,j}\vt^\b{}_i-
 X_{\a\b,i}\vt^\b{}_j\Big)\,.
 \end{equation}%\labe{KK-def}
 Thus, in fact, we have in (\ref{PreMax-x}), two first order
  partial differential equations
  for the entries of an  antisymmetric matrix $X_{\a\b}$.
Let us construct from this matrix  an antisymmetric  tensor $F_{ij}$
 \begin{equation}\label{F1-def}
 F_{ij}=X_{\mu\nu}\vt^\mu{}_i\vt^\nu{}_j\,,\qquad
X_{\mu\nu}=F_{ij}e_\mu{}^ie_\nu{}^j\,.
 \end{equation}%\labe{F1-def}
Substituting into (\ref{KK-def}), we derive
\begin{eqnarray}\label{KK1-def}
 K_{ijk}&=&F_{k[i,j]}-\frac 12 \, X_{\a\b}\Big[(\vt^\a{}_k\vt^\b{}_i)_{,j}-
 (\vt^\a{}_k\vt^\b{}_j)_{,i}\Big]\nonumber\\
 &=&F_{k[i,j]}-F_{km}C^m{}_{ij}-\frac 12 \left(F_{mi}\oG^m{}_{kj}-
 F_{mj}\oG^m{}_{ki}\right)\,.
 \end{eqnarray}%\labe{KK1-def}
Consequently, the first equation from (\ref{PreMax-x}) takes the
form
\begin{eqnarray}\label{PreMax1-x}
F_{[ij,k]}=\frac 23
(C^m{}_{ij}F_{km}+C^m{}_{jk}F_{im}+C^m{}_{ki}F_{jm})\,,
\end{eqnarray}%\labe{PreMax1-x}
while the second equation from (\ref{PreMax-x}) is rewritten as
\begin{eqnarray}\label{PreMax2-x}
F^i{}_{j,i}=-2F^i{}_mC^m{}_{ij}+
F_{kj}g^{ki}{}_{,i}+F_{mj}g^{ki}\oG^m{}_{ki}-F_{mi}g^{ki}\oG^m{}_{kj}\,.
\end{eqnarray}%\labe{PreMax2-x}
Observe first  a significant approximation to
(\ref{PreMax1-x}---\ref{PreMax2-x}). If  the right hand sides in
 both equations are
neglected, the equations take the form of the ordinary Maxwell
equations for the electromagnetic field in vacuum ---
 \begin{equation}\label{Max-Lor}
 F_{[ij,k]}=0\,, \qquad F^i{}_{j,i}=0\,.
 \end{equation}%\labe{Max-Lor}
In the coframe models,  the gravity is modeled by a variable
coframe field, i.e., by nonzero values of the quantities
$\oG_{ij}{}^k$. Consequently, the right hand sides of
(\ref{PreMax1-x}---\ref{PreMax2-x}) can be viewed as curved space
additions, i.e., as the  gravitational corrections to the
electromagnetic field equations.
In the flat spacetime, when a suitable coordinate system is chosen,
 these corrections are identically equal to zero. Consequently,
 in the flat spacetime, the invariance conditions
 (\ref{PreMax-x}) take the form of  the  vacuum Maxwell
 system.

On a curved manifold, the standard Maxwell equations are
formulated in a covariant form. Let us show that our
system
(\ref{PreMax1-x}---\ref{PreMax2-x}) is already covariant.
We rewrite (\ref{KK1-def}) as
\begin{eqnarray}\label{KK2-def}
 K_{ijk}=\frac 12(F_{ki,j}-F_{km}\oG{}^m{}_{ij}-
 F_{mi}\oG{}^m{}_{kj})-
 \frac 12 (\,\,i\longleftrightarrow j\,\,)\,.
 \end{eqnarray}%\labe{KK2-def}
Consequently,
\begin{eqnarray}\label{KK3-def}
 K_{ijk}=F_{k[i;j]}\,,
  \end{eqnarray}%\labe{KK3-def}
where the covariant derivative (denoted by the semicolon)
is taken relative to the Weitzenb\"{o}ck connection.
 Consequently, the system
(\ref{PreMax1-x}---\ref{PreMax2-x}) takes the covariant form
\begin{equation}\label{Max-Rie}
 F_{[ij;k]}=0\,, \qquad F^i{}_{j;i}=0\,.
 \end{equation}%\labe{Max-Rie}
 These equations are literally the same as the
 electromagnetic sector field equations of the Maxwell-Einstein system.
 The crucial difference is  encoded in the type of the covariant derivative.
 In the Maxwell-Einstein system, the covariant derivative is taken relative
 to the Levi-Civita connection, while, in our case,
 the corresponding connection is of   Weitzenb\"{o}ck.
 Observe that, due to our approach, the Weitzenb\"{o}ck connection is rather
natural in (\ref{Max-Rie}). Indeed, since the electromagnetic-type 
 field describes
 the local change of the coframe field, it should itself be referred only
 to the global changes of the coframe. As we have shown, such
 global transformations correspond precisely to the teleparallel
 geometry with the   Weitzenb\"{o}ck connections.

 \section{Geometrized coframe field model }
 \subsection{Generalized Einstein-Hilbert Lagrangian}
 One of the most important feature of the Einstein gravity theory is its pure geometrical content. The basic field variable of this theory is the metric tensor field $g_{ij}$. The action integral is given by the Einstein-Hilbert Lagrangian
\begin{equation}\label{EHGR}
^{\tt (GR)}{\cal A}=\int_M R\left(\LG^i_{jk}(g),g\right)*1\,,
\end{equation} 
where $R$ is the curvature scalar constructed from the metric tensor and its partial derivatives while $*1$ is the invariant volume element constructed from the metric tensor. 
When we restrict to the quasilinear  second order field equations the Lagrangian (\ref{EHGR}) is a unique possible. 

The coframe field model also constructed from the geometrical field variable --- coframe. Its Lagrangian however is taken as an arbitrary linear combination of the global $SO(1,3)$ invariants. The geometrical sense of this expression is  not clear. Although the coframe Lagrangian can be written in term of the torsion of the flat connection it does not mean that it corresponds to the Weitzenb\"{o}ck geometry with a flat curvature and a non-zero connection. Indeed also the standard Einstein-Hilbert Lagrangian  (\ref{EHGR}) can be rewritten in such a form. Moreover, as we have seen in the previous section, there is a wide class of connections all constructed from  Weitzenb\"{o}ck connection and its torsion. 
In particular, using the coframe Lagrangian in the form (\ref{3.2})  
we cannot answer the question: {\it What special geometry corresponds to the set of viable coframe models?}

Our proposal is to consider for the coframe Lagrangian an expression similar to (\ref{EHGR})
\begin{equation}\label{EHcof}
^{\tt (cof)}{\cal A}=\int_M R\Big(\G^i_{jk}(\vt^\a),g(\vt^\a)\Big)*1\,,
\end{equation}
which is constructed from the general free parametric coframe connection. Also the invariant volume element $*1$ is constructed here from the coframe field. Since the Levi-Civita connection is included as a special case of general coframe connection we have in (\ref{EHcof}) a generalization of the standard GR.

\subsection{Curvature of the coframe connection}
 %-------------------------------------

 \vspace{0.3cm}

\noindent{\bf{Riemannian curvature 2-form.}}
  %-------------------------------------
 We start with the definitions of the Riemannian curvature machinery.
 Although it is a classical subject of differential geometry \cite{kob1},
 in the case of a general connection of non-zero torsion and nonmetricity,
 slightly different notations are in use. Moreover, in this case,
  it is useful to apply the formalism of differential forms.
 We accept  the agreements used in metric-affine gravity \cite{Hehl:1994ue}.

 Let a connection 1-form $\GG_a{}^b$ referred to a general nonholonomic
 basis $(\theta^a,f_a)$ be given. The {\it curvature 2-form} is defined
 as
 \begin{equation}\label{CL1}
 \R_a{}^b=d\GG_a{}^b-\GG_a{}^c\wedge\GG_c{}^b\,.
 \end{equation}
 It satisfies two  fundamental identities:

 {\it  The first Bianchy  identity} involves the first order
 derivatives of the connection
 \begin{equation}\label{CL2}
 D\T^a-\R_b{}^a\wedge \theta^b=0\,, \qquad {\rm or}\qquad
 d\T^a+ \GG_b{}^a\wedge \T^b-\R_b{}^a\wedge \theta^b=0\,.
 \end{equation}

  {\it The second Bianchy  identity} involves the second
 order derivatives of the connection
 \begin{equation}\label{CL3}
 D\R_b{}^a=0\,,\qquad {\rm or}\qquad d\R_a{}^b+\GG_a{}^c
 \wedge \R_c{}^b-\GG_c{}^b\wedge \R_a{}^c=0\,.
 \end{equation}
It is useful to consider the Riemannian curvature of the coframe
connection to be referred to a basis composed from the elements
of the coframe field itself.  The
corresponded quantity
 \begin{equation}\label{CL3x}
 \R_\a{}^\b=d\GG_\a{}^\b-\GG_\a{}^\g\wedge\GG_\g{}^\b\,.
 \end{equation}
 is related to the generic  basis
 expression by the standard tensorial rule with the matrices of
 transformation $\vt^\a{}_if_a{}^i$
\begin{equation}\label{CL3xx}
 \R_a{}^b=\R_\a{}^\b(\vt^\a{}_if_a{}^i) (e_\b{}^j\theta^b{}_j)\,.
\end{equation}
From (\ref{CL3x}), we see that the Riemannian curvature of the
Weitzenb\"{o}ck connection is zero  being referred to a basis of
the coframe field. Due to (\ref{CL3x}), it is zero in an
arbitrary basis.

 Being referred to a coordinate basis, the Riemannian curvature 2-form reads
 \begin{eqnarray}\label{CL4-1}
 \R_i{}^j&=&d\GG_i{}^j-\GG_i^k\wedge\GG_k{}^j\\
 \label{CL4-2}&=&
 d\G^j{}_{in}\wedge dx^n-\G^k{}_{im}\G^j{}_{kn}dx^m\wedge dx^n\\
 \label{CL4-3}&=&
 \Big(\G^j{}_{in,m}-\G^k{}_{im}\G^j{}_{kn}\Big)dx^m\wedge dx^n\,.
 \end{eqnarray}
 The components of the Riemannian curvature 2-form
  \begin{equation}\label{CL5}
 \R_i{}^j=\frac 12\, R^j{}_{imn}dx^m\wedge dx^n
 \end{equation}
 are arranged in the  familiar expression of the {\it
 Riemannian curvature tensor}
  \begin{equation}\label{CL6}
 R^j{}_{imn}=\G^j{}_{in,m}-\G^j{}_{im,n}+\G^k{}_{in}\G^j{}_{km}
 -\G^k{}_{im}\G^j{}_{kn}\,.
 \end{equation}
 %-------------------------------------
 
 \vspace{0.3cm}

\noindent{\bf{Curvature scalar density.}}
  %----------------------------------------
 Curvature scalar plays an important role in physical applications.
 In fact, it is used as an integrand  in action of geometrical
field models --- Hilbert-Einstein Lagrangian density 
   \begin{equation}\label{SL1x}
 \L=R\,{\rm{vol}}=R*1\,,
\end{equation}
where star denotes the Hodge dual. In term of the curvature 2-form,
this expression is rewritten as
  \begin{equation}\label{SL1xx}
 \L= \R_{ij}\wedge*\,(dx^i\wedge dx^j)=\R_{\a\b}\wedge*\,\vt^{\a\b}\,.
 \end{equation}
where the abbreviation $\vt^{\a\b}=\vt^\a\wedge\vt^\b$ is used. 
Extracting in (\ref{SL1xx}) the total derivative term we obtain
   \begin{eqnarray}\label{SL2}
 \L&=&\left(d\GG_{\a\b}-\GG_\a{}^\g\wedge\GG_{\g\b}\right)
 \wedge *\,\vt^{\a\b}\nonumber\\&=&d\left(\GG_{\a\b}\wedge *\,\vt^{\a\b}
 \right)+
 %&&
 \GG_{\a\b}\wedge d*\vt^{\a\b}-\GG_\a{}^\g\wedge\GG_{\g\b}
 \wedge *\,\vt^{\a\b}\,.
 \end{eqnarray}
 For actual calculation of this quantity, it is useful to express  the connection
 1-form in the basis of the coframe field. We denote 
  \begin{equation}\label{SL3}
 \GG_{\a\b}=K_{\a\g\b}\vt^\g\,.
 \end{equation}
 Substituting it in the total derivative term of (\ref{SL2}) we
 have
  \begin{eqnarray}\label{SL4}
 d\left(\GG_{\a\b}\wedge*\,\vt^{\a\b}\right)&=&
 d\left(K_{\a\g\b}\vt^\g\wedge*\,\vt^{\a\b}\right)=
 (-1)^n d\left[K_{\a\g\b}*\left(e^\g\rfloor\,\vt^{\a\b}\right)\right]
 \nonumber\\&=&
 (-1)^n d\left[\left(K^\a{}_{\a\b}-K_{\b\a}{}^\a\right)*\vt^\b\right]\,.
\end{eqnarray}
The second term of (\ref{SL2}) reads
  \begin{equation}\label{SL5}
\GG_{\a\b}\wedge d*\vt^{\a\b}=K_{\a\g\b}\vt^\g\wedge
 d*\vt^{\a\b}%\nonumber\\
 %&=&
=K_{\a\g\b}\left[d\vt^\g\wedge *\,\vt^{\a\b}
 -d\left(\vt^\g\wedge *\,\vt^{\a\b}\right)\right]\,.
\end{equation}
Calculate:
 \begin{eqnarray}\label{SL6x}
 d\vt^\g\wedge *\,\vt^{\a\b}=\frac 12 C^\g{}_{\mu\nu}\vt^{\mu\nu}\wedge *\,\vt^{\a\b}=
 (-1)^{n+1}C^{\g\a\b}*1\,.
 \end{eqnarray}
 and
 \begin{eqnarray}\label{SL6xx}
 d\left(\vt^\g\wedge *\,\vt^{\a\b}\right)=(-1)^nd*(\eta^{\a\g}\vt^\b-\eta^{\b\g}\vt^\a)=
 (-1)^{n}\left(\eta^{\b\g}C^\a-\eta^{\a\g}C^\b\right)*1\,.
 \end{eqnarray}
Consequently the second term of (\ref{SL2}) takes the form
  \begin{eqnarray}\label{SL6}
\GG_{\a\b}\wedge d*(\vt^\a\wedge\vt^\b)&=&
 (-1)^n\left[K_{\a\g\b}C^{\g\a\b}-\left(K^\a{}_{\a\b}-K_{\b\a}{}^\a\right)C^\b\right]*1
 \end{eqnarray}
The third term of (\ref{SL2}) reads
  \begin{eqnarray}\label{SL7}
\GG_\a{}^\g\wedge\GG_{\g\b}\wedge*(\vt^\a\wedge\vt^\b)&=&
 K_{\a\mu}{}^\g K_{\g\nu\b}\vt^{\mu\nu}\wedge*\,\vt^{\a\b}
 \nonumber\\
 &=&(-1)^n\left(K^{\a\b\g}K_{\g\a\b}-K^\a{}_{\a\g}K^{\g\b}{}_\b\right)*1\,.
 \end{eqnarray}
 Consequently the  Lagrangian density takes the form
  \begin{eqnarray}\label{SL8}
 \L(-1)^n&=&d\left[\left(K^\a{}_{\a\b}-K_{\b\a}{}^\a\right)*\vt^\b\right]+
\left[K_{\a\g\b}C^{\g\a\b}-\left(K^\a{}_{\a\b}-K_{\b\a}{}^\a\right)C^\b\right]*1
\nonumber\\
&&-\left(K^{\a\b\g}K_{\g\a\b}-K^\a{}_{\a\g}K^{\g\b}{}_\b\right)*1
\,.
  \end{eqnarray}
 Due to (\ref{mcc12}), the tensor $K_{\a\g\b}$ is of the form
 \begin{eqnarray}\label{SL9}
 K_{\a\g\b}=\a_1C_{\b\g\a}+\a_2C_\g\eta_{\a\b}
+\a_3C_\a\eta_{\b\g}+
\b_1C_\b\eta_{\a\g}+\b_2C_{\g\a\b}+\b_3C_{\a\b\g}\,,
\end{eqnarray}
 Substituting this expression in (\ref{SL8}) we obtain a total
 derivative term plus a sum of terms which are quadratic in $C_{\a\b\g}$.
 Since (\ref{3.2}) is the most general expression quadratic in
 $C_{\a\b\g}$, the following statement is clear. 

 \vspace{0.3cm} \noindent {\bf Proposition 7}
\noindent  {\it The Hilbert-Einstein Lagrangian of the
 general metric-coframe connection (\ref{mcc12}) is equivalent up to a total
 derivative term to the general coframe Lagrangian
 \begin{equation}\label{SL10}
 R(\GG_{\a\b})*1=\zeta_0 d(C_\a*\vt^\a)+\left(\zeta_1C_{\a\b\g}C^{\a\b\g}+
\zeta_2C_{\a\b\g}C^{\b\g\a} +\zeta_3C_\a C^\a\right)*1\,,
 \end{equation}
 where the parameters $\zeta_i$ are expressed by second order polynomials of the coefficients $\a_i,\b_i$.
 }
 \vspace{0.3cm}
 
 The actual expressions for the coefficients $\zeta_i$ are rather involved. We discuss the parameter $\zeta_0$ in sequel.  
 % \begin{eqnarray}\label{SL11}
% \zeta_0&=&-\a_1-n(\a_3-\b_1)+2\b_2-\b_3\,,\\
% \label{SL12}
% \zeta_1&=&\b_2+\b_3\\
 %\label{SL13}
 %\zeta_2&=&\a_1\\
% \label{SL14}
 %\zeta_3&=&\b_1-\a_3-\zeta_0+\big(a_2+a_3+(n+1)\b_1+\b_2-\b_3\big)\cdot 
%%  \end{eqnarray}}
  \subsection{Einstein-Hilbert Lagrangian without second order derivatives}
 %---------------------------
 It is well known that in GR the Einstein-Hilbert Lagrangian involves the second order derivatives of the metric tensor. These terms joint in a total derivative term which is not relevant for the field equation. Although, the total derivative terms cannot consistently dropped out. In particular, the quantization procedure requires an addition of a boundary term in order to compensate the total derivative \cite{York:1972sj}, \cite{Gibbons:1976ue}. Let us calculate the  total derivative term in our model. 
 Withe (\ref{SL9}) we have 
 \begin{equation}\label{tot1}
 K^\a{}_{\a\b}=\eta^{\a\g}K_{\a\g\b}=\left[\a_2+\a_3+(n+1)\b_1+\b_2-\b_3\right] C_\b\,,
 \end{equation}
 and
 \begin{equation}\label{tot2}
 K_{\b\a}{}^\a=\eta^{\b\g}K_{\a\g\b}=\left[\a_1+\a_2+(n+1)\b_1+\b_2-\b_3\right] C_\b\,.
 \end{equation}
 Thus
\begin{equation}\label{tot3}
 d\left[\left(K^\a{}_{\a\b}-K_{\b\a}{}^\a\right)*\vt^\b\right] =
-\left[\a_1+n(\a_3-\b_1)+2\b_2-\b_3\right] d \left(C_\b*\vt^\b\right)\,.
 \end{equation}
 Consequently, the coefficient $\zeta_0$ in (\ref{SL10}) takes the form
 \begin{equation}\label{tot4}
 \zeta_0=\a_1+n(\a_3-\b_1)+2\b_2-\b_3\,.
 \end{equation}
 For the Weitzenb\"{o}ck connection, this coefficient is zero together with all other terms of the Lagrangian.  For the Levi-Civita connection,  $\zeta_0= -2$ on a manifold of an 
arbitrary dimension. 

We can identify now  a family of coframe connections without a total derivative term at all. 
It is enough to require 
\begin{equation}\label{tot5}
 \a_1+n(\a_3-\b_1)+2\b_2-\b_3=0\,.
 \end{equation}
 The corresponding connection is given by
 \begin{eqnarray}\label{tot6}
\G_{ijk}&=&\,\oG_{ijk}+\a_1(C_{ijk}+C_{kij})+\a_2g_{ik}C_j+\a_3(g_{ij}C_k+nC_{kij})+
\nonumber\\
&&\b_1(g_{jk}C_i-nC_{kij})+\b_2(C_{jki}+2C_{kij})\,.
\end{eqnarray}
 This family includes the metric-compatible connections 
 \begin{equation}\label{tot7}
 \G_{ijk}=\oG_{ijk}+\a_1(C_{ijk}+C_{jki}+3C_{kij})+\a_2(g_{ik}C_j-g_{jk}C_i+nC_{kij})\,,
 \end{equation}
and the symmetric (torsion-free) connections
\begin{equation}\label{tot8}
 \G_{ijk}=\,\oG_{i(jk)}+\a_2\left(g_{ik}C_j+g_{ij}C_k\right)+
\b_1g_{jk}C_i+[1-n(\a_2-\b_1)]\left(C_{jki}+C_{kji}\right)\,.
 \end{equation}
Also the gauge invariant connections (\ref{gt13}) can be found into the family (\ref{tot6}).

Consequently we identified a remarkable property of the coframe geometry. 
There is a family of coframe connections which standard Einstein-Hilbert Lagrangian does not involve second order derivatives terms at all. It means that there is a family of coframe models with a geometrical  Lagrangian which is completely equivalent to the Yang-Mills Lagrangians of particle physics. 
 
 \section{Conclusion}
 %--------------------------------------
 GR is a well-posed classical field theory for 10 independent variables --- the components of the metric tensor. Although, this theory is completely satisfactory in the pure gravity sector, its possible extensions to other physics phenomena is rather problematic. In particular, the description of fermions on a curved space and the supergravity constructions require a richer set of 16 independent variables. These variables can be assembled in a coframe field, i.e., a local set of four linearly independent 1-forms. Moreover, in supergravity, it is necessary to involve a special flat connection constructed from the derivatives of the coframe field. These facts justify the study of the field models based on a coframe variable alone.  
 
 The classical field construction of the coframe gravity is based on a Yang-Mills-type Lagrangian which is a linear combination of quadratic terms with dimensionless coefficients. Such model turns to be satisfactory in the gravity sector and  has the viable Schwarzschild solutions even being alternative to the standard GR. Moreover, the coframe model treating of the gravity energy makes it even preferable than the ordinary GR where the gravity energy cannot be defined at all. A principle problem that the coframe gravity construction does not have any connection to a specific geometry even being constructed from the geometrical meaningful objects. A geometrization of the coframe gravity is an aim of this chapter.

 We construct a general family of coframe connections which involves as the special cases  the  Levi-Civita connection of GR and  the flat Weitzenb\"{o}ck connection.  Every specific connection generates a geometry of a specific type. We identify the subclasses of  metric-compatible and  torsion-free connections. Moreover we study the local linear  transformations of the coframe fields and identify a class of connections which are invariant under restricted coframe transformations. Quite remarkable that the restriction conditions are necessary approximated by a Maxwell-type system of equations.    

On a basis of the coframe geometry, we propose a geometric action for the coframe gravity. It has the same form as the Einstein-Hilbert action of GR, but the scalar curvature is constructed from the general coframe connection. We show that this geometric Lagrangian is equivalent to the coframe Lagrangian up to a total derivative term. Moreover there is a family of coframe connections which Lagrangian does not include the higher order terms. In this case, the equivalence is complete. 

 However,  the Hilbert-Einstein-type action  itself is not enough to predict a unique coframe connection. Indeed, the coframe connection has six free parameters, while the action involves only four of their combinations. Moreover, one combination represents a total derivative term in Lagrangian which does not influence the field equations. So the gravity action itself is not defined uniquely the geometry on the base  manifold. It should not be, however,  a problem. Indeed, the gravitational field is not a unique physical field. Moreover, gravity  does not even exist without matter fields as its origin. An action for an arbitrary (non-scalar) field necessary  involves the connection. So the problem can be formulated as following: To find out which matter field has to be added to the coframe Lagrangian in order to predict uniquely the type of the coframe  connection and consequently the geometry of the  underlying manifold. This problem can serve as a basis for  future investigation.
  
\section*{Acknowledgements}  I thank Shmuel Kaniel, Yakov Bekenstein, Yaakov Friedman (Jerusalem), Friedrich W. Hehl (Cologne and Missouri-Columbia), Yuri N. Obukhov (Moscow and Cologne), and Roman Jackiw (MIT) for fruitful discussions of the coframe gravity.

\renewcommand{\theequation}{A.\arabic{equation}}
  % redefine the command that creates the equation no.
  \setcounter{equation}{0}  % reset counter 
  \section{Appendix --- differential form notations}  % use *-form to suppress numbering

% \vspace{0.3cm}

%\appendix{Appendix --- differential form notations }

%\section*{Appendix --- differential form notations }
%-------------------------------------------------------
We collect here some algebraic rules which are useful for
calculations with the differential forms. Recall that we are
working on an $n+1$ dimensional manifold.

{\it 1. Interior product}\\
 In a basis of 1-forms $\vt^\a$, a $p$-form $\Psi$
 is expressed as
\begin{equation}\label{a0-2x}
 \Psi=\frac 1{p!} \Psi_{\a_1\cdots\a_p}
 \vt^{\a_1}\wedge\cdots\wedge\vt^{\a_p}\,.
  \end{equation}
 Interior product couples the basis vectors and basis 1-forms as
  \begin{equation}\label{a0-0}
 e_\a\rfloor \vt^\b=\d^b_\a\,.
 \end{equation}
 By bilinearity and the Leibniz-type rule,
  \begin{equation}\label{a0-1}
e_\a\rfloor (w_1\wedge w_2)=(e_\a\rfloor w_1)\wedge w_2+
 (-1)^{{\rm deg} w_1}w_1\wedge (e_\a\rfloor w_2)\,,
\end{equation}
 the definition of the interior product is extended  to
 forms of arbitrary degree.
 Mixed applications of the exterior and interior products to a $p$-form $w$
 satisfy the relations
    \begin{equation}\label{a0-4}
 \vt^\a\wedge(e_a\rfloor w)=pw\,,
 \end{equation}
 and
   \begin{equation}\label{a0-5}
 e_a\rfloor (\vt^\a\wedge w)=(n-p)w\,.
 \end{equation}
 %--------------
{\it 2. Hodge star operator}\\
  %--------------
 The Hodge star operator maps $p$-forms into $(n+1-p)$-forms.
 In  a pseudo-orthonormal basis $\vt^\a$,  the metric tensor
 is represented by the constant components
 $\eta_{\a\b}={\rm diag}(-1,1,\cdots,1)$.
 In this case, the Hodge star operator is defined as
  \begin{equation}\label{a0-2xy}
 *\Psi=\frac 1{p!(n+1-p)!}
 \Psi_{\a_0\cdots\a_p}\eta^{\a_0\b_0}\cdots\eta^{\a_p\b_p}
 \varepsilon_{\b_0\cdots\b_n}
 \vt^{\b_{p+1}\wedge\cdots\wedge\vt^{\b_n}}\,,
  \end{equation}
 where the permutation symbol is normalized as
 \begin{equation}\label{a0-2xx}
 \varepsilon_{0\cdots n}=1\,,\qquad \varepsilon^{0\cdots n}=-1\,.
  \end{equation}
 For the basis forms themselves, this formula can be rewritten as
  \begin{equation}\label{a0-3x}
 *(\vt_{\a_0}\wedge\cdots\wedge\vt_{\a_p})=\frac 1{(n+1-p)!}
 \varepsilon_{\a_0\cdots\a_p\b_1\cdots\b_{n-p}}
 \vt^{\b_1}\wedge\cdots\wedge\vt^{\b_{n-p}}\,.
 \end{equation}
 In particular,
   \begin{equation}\label{a0-3xx}
 *(\vt_{\a_0}\wedge\cdots\wedge\vt_{\a_n})=\varepsilon_{\a_0\cdots\a_n}\,,
 \qquad
 *1=\frac 1{n!}\varepsilon_{\a_0\cdots\a_n}
 \vt^{\a_1}\wedge\cdots\wedge\vt^{\a_n}\,.
 \end{equation}
  When the Hodge map defined by a Lorentzian-type metric $\eta_{\a\b}$ it acts on a $p$-form $w$
  \begin{equation}\label{a0-2}
 **w=(-1)^{p(n+1-p)+1}w=(-1)^{pn+1}w\,.
 \end{equation}
 For the forms $w_1,w_2$ of the same degree,
   \begin{equation}\label{a0-3}
 w_1\wedge * w_2=w_2\wedge *w_1\,.
 \end{equation}
 With the Hodge map, the wedge product can be transformed into the interior product
 and vice versa  by the relations
  \begin{equation}\label{a0-6}
*(w\wedge\vt_\a)=e_\a\rfloor *w\,,
 \end{equation}
 and
  \begin{equation}\label{a0-7}
 \vt_\a\wedge *w =(-1)^{n(n-p)}*(e_\a\rfloor w)\,.
  \end{equation}
 %----------------------
 {\it{3. Exterior derivative and coderivative of the coframe
 field}}\\
  %----------------------
 We express  the exterior derivative of the  coframe field as
  \begin{equation}\label{a0-8}
 d\vt^\a=\frac 12 C^\a{}_{\b\g}\vt^\b\wedge\vt^\g\qquad C_\a=C^\mu{}_{\mu\a}\,.
 \end{equation}
 The divergence of the coframe 1-form is
\begin{equation}\label{a0-9}
d*\vt_\a=-C_{\a}*1\,.
\end{equation}
Indeed, using (\ref{a0-3x}) we calculate
 \begin{eqnarray}\label{a0-10x}
 d*\vt_\a&=&\frac 1{n!}\varepsilon_{\a\b_1\cdots\b_{n}}d(\vt^{\b_1}
 \wedge\cdots\wedge\vt^{\b_{n}})\nonumber\\
 %&=&\frac 1{2n!}\varepsilon_{\a\b_1\cdots\b_{n}}\Big(C^{\b_1}{}_{\mu\nu}
% \vt^\mu\wedge\vt^\nu\wedge\vt^{\b_2}\wedge\cdots\wedge\vt^{\b_{n}}-
 %C^{\b_2}{}_{\mu\nu}
 %%\cdots\Big)
 %\nonumber\\
 &=&\frac 1{2(n-1)!}\varepsilon_{\a\b_1\cdots\b_{n}}C^{\b_1}{}_{\mu\nu}
 \vt^\mu\wedge\vt^\nu\wedge\vt^{\b_2}\wedge\cdots\wedge\vt^{\b_{n}}\,.
 \end{eqnarray}
 Using (\ref{a0-3xx}) and (\ref{a0-2}) we have
 \begin{equation}\label{a0-10}
 \vt^\mu\wedge\vt^\nu\wedge\vt^{\b_2}\wedge\cdots\wedge\vt^{\b_{n}}=
 -\varepsilon^{\mu\nu\b_2\cdots\b_{n}}*1\,.
 \end{equation}
 Consequently,
  \begin{eqnarray}\label{a0-10xx}
 d*\vt_\a&=&-\frac 1{2(n-2)!}\varepsilon_{\a\b_1\cdots\b_{n-1}}
\varepsilon^{\mu\nu\b_2\cdots\b_{n-1}}
 C^{\b_1}{}_{\mu\nu}*1=\nonumber\\&=&
 \frac 12 (\d^\mu_\a\d^\nu_{\b_1}-\d^\nu_\a\d^\mu_{\b_1})
 C^{\b_1}{}_{\mu\nu}*1=C^\mu{}_{\a\mu}*1=-C_{\a}*1\,.
 \end{eqnarray}

In a coordinate basis we consider the tensors
 \begin{equation}\label{Diform1}
 C^i{}_{jk}=\frac12 \left(\oG^i{}_{jk}-\oG^i{}_{kj}\right)\,,\qquad C_i=C^m{}_{mi}\,.
 \end{equation}
 It is easy to check the relations
 \begin{equation}\label{Diform1x}
 C^i{}_{jk}=C^\a{}_{\b\g}e_\a{}^i\vt^\b{}_j\vt^\g{}_k\,,\qquad C_i=C_\a\vt^\a{}_i\,.
 \end{equation}
  Define a
non-indexed (scalar-valued) 1-form
\begin{equation}\label{Diform2}
\A=e_\mu\rfloor d\vt^\mu=2\vt^\mu{}_{[i,j]}e_\mu{}^i\,dx^j=2C_idx^i=3C_\a\vt^\a\,.
\end{equation}
On a manifold with a metric $ g=\eta_{\mu\nu}\vt^\mu\otimes\vt^\nu$
 (Section 3), we
define, in addition,  a scalar-valued 3-form
\begin{eqnarray}\label{Diform3}
\B&=&\eta_{\mu\nu}\,d\vt^\mu\wedge\vt^\nu=
-\eta_{\mu\nu}\vt^\mu{}_{i,j}\vt^\nu{}_k\,dx^i\wedge dx^j\wedge
dx^k\nonumber\\
&=&C_{ijk}dx^i\wedge dx^j\wedge dx^k=C_{\a\b\g}\vt^\a\wedge\vt^\b\wedge\vt^\g\,.
\end{eqnarray}

The operations of symmetrization and antisymmetrization of tensors are used
 here  in the normalized form:
\begin{equation}\label{Diform4}
(a_1\cdots a_p)=\frac 1{p!}\,{\rm Sym}(a_1\cdots a_p)\,,\qquad [a_1\cdots
a_p]=\frac 1{p!}{\rm Ant}(a_1\cdots a_p)\,.
\end{equation}


\addcontentsline{toc}{section}{\refname}

\begin{thebibliography}{99}

%%%%%%%%%%%%%%%%%%%   MAG


%\cite{Hehl:1976kj}
\bibitem{Hehl:1976kj}
  F.~W.~Hehl, P.~Von Der Heyde, G.~D.~Kerlick and J.~M.~Nester,
  %``General Relativity With Spin And Torsion: Foundations And Prospects,''
  Rev.\ Mod.\ Phys.\  {\bf 48}, 393 (1976).
  
   %\cite{Hojman:1978yz}
\bibitem{Hojman:1978yz}
  S.~Hojman, M.~Rosenbaum, M.~P.~Ryan and L.~C.~Shepley,
  %``Gauge Invariance, Minimal Coupling, And Torsion,''
  Phys.\ Rev.\ D {\bf 17}, 3141 (1978).
  %%CITATION = PHRVA,D17,3141;%%
  
  %\cite{Kopczynski:1979ee}
\bibitem{Kopczynski:1979ee}
  W.~Kopczynski,
  %``Metric Affine Unification Of Gravity And Gauge Theories,''
  Acta Phys.\ Polon.\  B {\bf 10} (1979) 365.
  
  \bibitem{McCrea:1992wa}
  J.~D.~McCrea,
  %``Irreducible decompositions of non-metricity, torsion, curvature and Bianchi
  %identities in metric-affine spacetimes,''
  Class.\ Quant.\ Grav.\  {\bf 9}, 553 (1992).
  
 %\cite{Hehl:1994ue}
\bibitem{Hehl:1994ue}
F.~W.~Hehl, J.~D.~McCrea, E.~W.~Mielke and Y.~Neeman,
%``Metric affine gauge theory of gravity: Field equations, Noether identities,
%world spinors, and breaking of dilation invariance,''
Phys.\ Rept.\  {\bf 258}, 1 (1995);
%[arXiv:gr-qc/9402012].

%\cite{Gronwald:1995em}
\bibitem{Gronwald:1995em}
  F.~Gronwald and F.~W.~Hehl,
  %``On the gauge aspects of gravity,''
  arXiv:gr-qc/9602013.
  
  %\cite{Giachetta:1995bj}
\bibitem{Giachetta:1995bj}
  G.~Giachetta and G.~Sardanashvily,
  %``Stress-Energy-Momentum of Affine-Metric Gravity. Generalized Komar
  %Superportential,''
  Class.\ Quant.\ Grav.\  {\bf 13}, L67 (1996)
  
%\cite{Obukhov:1996ka}
\bibitem{Obukhov:1996ka}
  Yu.~N.~Obukhov, E.~J.~Vlachynsky, W.~Esser and F.~W.~Hehl,
  %``Effective Einstein theory from metric affine gravity models via irreducible
  %decompositions,''
  Phys.\ Rev.\  D {\bf 56}, 7769 (1997).
  
  %\cite{Socorro:1998fm}
\bibitem{Socorro:1998fm}
  J.~Socorro, C.~Lammerzahl, A.~Macias and E.~W.~Mielke,
  %``Multipole solutions in metric-affine gravity,''
  Phys.\ Lett.\  A {\bf 244}, 317 (1998)
  
  %\cite{Garcia:1998jw}
\bibitem{Garcia:1998jw}
  A.~Garcia, F.~W.~Hehl, C.~Laemmerzahl, A.~Macias and J.~Socorro,
  %``Pleba\'nski-Demia\'nski-like solutions in metric-affine gravity,''
  Class.\ Quant.\ Grav.\  {\bf 15}, 1793 (1998)
  
  %\cite{Gronwald:1997bx}
\bibitem{Gronwald:1997bx}
  F.~Gronwald,
  %``Metric-affine gauge theory of gravity. I: Fundamental structure and  field
  %equations,''
  Int.\ J.\ Mod.\ Phys.\  D {\bf 6}, 263 (1997)
  
  %\cite{Hehl:1999sb}
\bibitem{Hehl:1999sb}
  F.~W.~Hehl and A.~Macias,
  %``Metric-affine gauge theory of gravity. II: Exact solutions,''
  Int.\ J.\ Mod.\ Phys.\  D {\bf 8}, 399 (1999)
  %[arXiv:gr-qc/9902076].
  
  %\cite{Puetzfeld:2001ur}
\bibitem{Puetzfeld:2001ur}
  D.~Puetzfeld,
  %``A cosmological model in Weyl-Cartan spacetime. I: Field equations and
  %solutions,''
  Class.\ Quant.\ Grav.\  {\bf 19}, 3263 (2002)
  
  %\cite{Godina:2000ey}
\bibitem{Godina:2000ey}
  M.~Godina, P.~Matteucci and J.~A.~Vickers,
  %``Metric-affine gravity and the Nester-Witten 2-form,''
  J.\ Geom.\ Phys.\  {\bf 39}, 265 (2001)
  
  %\cite{Vassiliev:2003dk}
\bibitem{Vassiliev:2003dk}
  D.~Vassiliev,
  %``Quadratic metric-affine gravity,''
  Annalen Phys.\  {\bf 14}, 231 (2005)
  
  %\cite{Pasic:2005qr}
\bibitem{Pasic:2005qr}
  V.~Pasic and D.~Vassiliev,
  %``PP-waves with torsion and metric-affine gravity,''
  Class.\ Quant.\ Grav.\  {\bf 22}, 3961 (2005)
  
  %\cite{Obukhov:2006gy}
\bibitem{Obukhov:2006gy}
  Y.~N.~Obukhov,
  %``Plane waves in metric-affine gravity,''
  Phys.\ Rev.\  D {\bf 73}, 024025 (2006)
  %[arXiv:gr-qc/0601074].
  %%%%%%%%%%%%%%%  Coframe gravity
  
  
\bibitem{cit1}
 %\cite{Goenner:2004se}
%\bibitem{Goenner:2004se}
H.~F.~M.~Goenner,
%``On the history of unified field theories,''
Living Rev.\ Rel.\  {\bf 7}, 2 (2004).



\bibitem{Hayash:1979}
%
 K.~Hayashi and T.~Shirafuji,
%``New General Relativity,''
Phys.\ Rev.\ D {\bf 19}, 3524 (1979)
%[Addendum-ibid.\ D {\bf 24}, 3312 (1982)];




  \bibitem{Sezgin:1979zf}
  E.~Sezgin and P.~van Nieuwenhuizen,
  %``New Ghost Free Gravity Lagrangians With Propagating Torsion,''
  Phys.\ Rev.\  D {\bf 21}, 3269 (1980).

\bibitem{Nitsch:1979qn}
%
 J.~Nitsch and F.~W.~Hehl,
%``Translational Gauge Theory Of Gravity: Postnewtonian Approximation
%And Spin
%Precession,''
Phys.\ Lett.\ B {\bf 90}, 98 (1980);
%
\bibitem{Mueller-Hoissen:1983vc}
F.~Mueller-Hoissen and J.~Nitsch,
%``Teleparallelism - A Viable Theory Of Gravity?,''
Phys.\ Rev.\ D {\bf 28}, 718 (1983);

  
  \bibitem{Kuhfuss:1986rb}
  R.~Kuhfuss and J.~Nitsch,
  %``PROPAGATING MODES IN GAUGE FIELD THEORIES OF GRAVITY,''
  Gen.\ Rel.\ Grav.\  {\bf 18}, 1207 (1986).
%
\bibitem{Mielke:1992te}
E.~W.~Mielke,
%``Ashtekar's complex variables in general
%relativity and its teleparallelism
%equivalent,''
Annals Phys.\  {\bf 219}, 78 (1992);
 %
\bibitem{Muench:1998ay}
U.~Muench, F.~Gronwald and F.~W.~Hehl,
%``A brief guide to variations in teleparallel
%gauge theories of gravity  and
%the Kaniel-Itin model,''
Gen.\ Rel.\ Grav.\  {\bf 30}, 933 (1998);
 %
\bibitem{Tung:1998dw}
R.~S.~Tung and J.~M.~Nester,
%``The quadratic spinor Lagrangian is equivalent to
%the teleparallel theory,''
Phys.\ Rev.\ D {\bf 60}, 021501 (1999);
%[arXiv:gr-qc/9809030].
%

\bibitem{Itin:1999zs}
Y.~Itin and S.~Kaniel,
%``On a class of invariant coframe operators with
%application to gravity,''
J.\ Math.\ Phys.\  {\bf 41}, 6318 (2000)
%[arXiv:gr-qc/9907023];

 
 %
\bibitem{Blagojevic:2000pi}
M.~Blagojevic and M.~Vasilic,
%``Gauge symmetries of the teleparallel theory of gravity,''
Class.\ Quant.\ Grav.\  {\bf 17}, 3785 (2000);

%\cite{Itin:1999wi}
\bibitem{Itin:1999wi}
Y.~Itin,
%``Coframe teleparallel models of gravity. Exact solutions,''
Int.\ J.\ Mod.\ Phys.\ D {\bf 10}, 547 (2001)
%[arXiv:gr-qc/9912013];
%\cite{Blagojevic:2000qs}
\bibitem{Blagojevic:2000qs}
  M.~Blagojevic and I.~A.~Nikolic,
  %``Hamiltonian structure of the teleparallel formulation of GR,''
  Phys.\ Rev.\  D {\bf 62}, 024021 (2000)
  
  \bibitem{Shapiro:2001rz}
  I.~L.~Shapiro,
  %``Physical aspects of the space-time torsion,''
  Phys.\ Rept.\  {\bf 357}, 113 (2001)
  %[arXiv:hep-th/0103093].

\bibitem{Hammond:2002rm}
  R.~T.~Hammond,
  %``Torsion gravity,''
  Rept.\ Prog.\ Phys.\  {\bf 65}, 599 (2002).
%\cite{Itin:2001bp}
\bibitem{Itin:2001bp}
Y.~Itin,
%``Energy-momentum current for coframe gravity,''
Class.\ Quant.\ Grav.\  {\bf 19}, 173 (2002);
%[arXiv:gr-qc/0111036].
\bibitem{Itin:2001xz}
Y.~Itin,
%``Coframe energy-momentum current. Algebraic properties,''
Gen.\ Rel.\ Grav.\  {\bf 34}, 1819 (2002);


\bibitem{Obukhov:2002tm}
Y.~N.~Obukhov and J.~G.~Pereira,
%``Metric-affine approach to teleparallel gravity,''
Phys.\ Rev.\ D {\bf 67}, 044016 (2003)
%[arXiv:gr-qc/0212080];

 \bibitem{Itin:2003jp}
Y.~Itin,
%``Noether currents and charges for Maxwell-like Lagrangians,''
J.\ Phys.\ A {\bf 36}, 8867 (2003)
%[arXiv:math-ph/0307003].
%\cite{Leclerc:2004uu}
\bibitem{Leclerc:2004uu}
  M.~Leclerc,
  %``On the teleparallel limit of Poincare gauge theory,''
  Phys.\ Rev.\  D {\bf 71}, 027503 (2005)
  %[arXiv:gr-qc/0411119].
  
  %\cite{Obukhov:2006fy}
\bibitem{Obukhov:2006fy}
  Y.~N.~Obukhov, G.~F.~Rubilar and J.~G.~Pereira,
  %``Conserved currents in gravitational models with quasi-invariant
  %Lagrangians: Application to teleparallel gravity,''
  Phys.\ Rev.\  D {\bf 74}, 104007 (2006)
 % [arXiv:gr-qc/0610092].
  
\bibitem{Itin:2004ig} Y.~Itin, J. Math. Phys.  {\bf 46}  12501 (2005).

\bibitem{Estabrook:2005ms}
  F.~B.~Estabrook,
  %``Conservation laws for vacuum tetrad gravity,''
  Class.\ Quant.\ Grav.\  {\bf 23}, 2841 (2006)
  
%\cite{Itin:2006pd}
\bibitem{Itin:2006pd}
  Y.~Itin,
  %``Maxwell-Type Behaviour From A Geometrical Structure,''
  Class.\ Quant.\ Grav.\  {\bf 23} (2006) 3361.
  
%\cite{Nashed:2007sc}
\bibitem{Nashed:2007sc}
  G.~G.~L.~Nashed and T.~Shirafuji,
  %``Reissner-Nordstroem spacetime in the tetrad theory of gravitation,''
  Int.\ J.\ Mod.\ Phys.\  D {\bf 16}, 65 (2007)
%%%%%% Coframe used
  
  

 \bibitem{Ashtekar:1987gu}
  A.~Ashtekar,
  %``New Hamiltonian Formulation Of General Relativity,''
  Phys.\ Rev.\ D {\bf 36}, 1587 (1987).

 \bibitem{Deser:1976ay}
      S.~Deser and C.~J.~Isham,
      %``Canonical Vierbein Form Of General Relativity,''
      Phys.\ Rev.\ D {\bf 14}, 2505 (1976).
 
 \bibitem{Nester:1994du}
  J.~M.~Nester and R.~S.~Tung,
  %``Another positivity proof and gravitational energy localizations,''
  Phys.\ Rev.\ D {\bf 49}, 3958 (1994)
 % [arXiv:gr-qc/9401002].

 \bibitem{Deser:1974cy}
  S.~Deser and P.~van Nieuwenhuizen,
  %``Nonrenormalizability Of The Quantized Dirac - Einstein System,''
  Phys.\ Rev.\ D {\bf 10}, 411 (1974).

 \bibitem{spin}
  P.~G.~Bergmann, V.~de Sabbata, G.~T.~Gillies and P.~I.~Pronin,
{\it International School of Cosmology and Gravitation: 15th Course: Spin in Gravity: Is it Possible to Give an Experimental Basis
to Torsion?}, Erice, Italy, 13-20 May 1997.


 \bibitem{VanNieuwenhuizen:1981ae}
  P.~Van Nieuwenhuizen,
  %``Supergravity,''
  Phys.\ Rept.\  {\bf 68}, 189 (1981).

 
\bibitem{Perez:15jz}
  A.~Perez and C.~Rovelli,
  %``Physical Effects Of The Immirzi Parameter In Loop Quantum Gravity,''
  Phys.\ Rev.\ D {\bf 73}, 044013 (2006)
  
  

\bibitem{kob1} S. Kobayashi and K. Nomizu, {\it Foundations of Differential Geometry}, vol. 1 and 2, Interscience Tracts in Pure and Applied Mathematics, Interscience Publ., New-York, 1969.

\bibitem{kob2} S. Kobayashi, {\it Transformation Groups in Differential Geometry}, Springer-Verlag, 1972.

\bibitem{Thomas} T.Y. Thomas: {\it The differential invariants on generalized spaces}, Cambridge, The University Press, 1934.

\bibitem{Schou} J.A. Schouten, {\it Ricci-Calculus,
An Introduction to Tensor Analysis and its Geometrical Applications}
(2nd ed.,
         Springer-Verlag, New York, 1954).
%\bibitem{Wald-book}
%  R.~M.~Wald,
 % ``General Relativity,''
%\href{http://www.slac.stanford.edu/spires/find/hep/www?irn=1334239}{SPIRES entry}
%{\it  Chicago, Usa: Univ. Pr. ( 1984) 491p}

\bibitem{Wald2} V. Iyer, R.M. Wald \emph{Phys.Rev.} , \textbf{D50}, (1994),  
846-864.


\bibitem{gyros} F.W.~Hehl and Yu.N.~Obukhov,
 %{\em How does the
 %   electromagnetic field couple to gravity, in particular to metric,
 %   nonmetricity, torsion, and curvature?} In: {\em Gyros, Clocks,
 %   Interferometers\ldots: Testing Relativistic Gravity in Space.} C.\
 % L\"ammerzahl et al., eds.
 Lecture Notes in Physics Vol.~{\bf 562}
  (Springer: Berlin, 2001) pp.\ 479-504. %; arXiv.org/gr-qc/0001010.


%\cite{York:1972sj}
\bibitem{York:1972sj}
  J.~W.~.~York,
  %``Role of conformal three geometry in the dynamics of gravitation,''
  Phys.\ Rev.\ Lett.\  {\bf 28}, 1082 (1972).
  
%\cite{Gibbons:1976ue}
\bibitem{Gibbons:1976ue}
  G.~W.~Gibbons and S.~W.~Hawking,
  %``Action Integrals And Partition Functions In Quantum Gravity,''
  Phys.\ Rev.\  D {\bf 15}, 2752 (1977).
  
  
\end{thebibliography}
\end{document}